\documentclass{article}

\usepackage{arxiv}

\usepackage{microtype}      

\usepackage[utf8]{inputenc} 
\usepackage[T1]{fontenc}    
\usepackage[english]{babel}

\usepackage{dsfont}
\usepackage{amsthm,amssymb,bm}
\usepackage{amsfonts}       
\usepackage{amsmath,etoolbox}
\usepackage{nicefrac}       
\usepackage{pifont}
\usepackage{mathtools}

\usepackage{chemarrow} 
\usepackage{graphicx}
\usepackage{svg}
\usepackage{subfig}
\usepackage[section]{placeins} 

\usepackage{booktabs}
\usepackage{adjustbox}
\usepackage{multirow}

\usepackage{makecell}

\usepackage{todonotes}

\usepackage[ruled]{algorithm2e}

\usepackage{enumitem}

\usepackage{hyperref}       
\usepackage{url}            

\usepackage{natbib}
\usepackage{doi}

\usepackage[toc,title,page]{appendix}

\usepackage{float}

\title{De novo peptide sequencing rescoring and FDR estimation with Winnow}

\author
    {\href{https://orcid.org/0009-0009-7514-677X}{\includegraphics[scale=0.06]{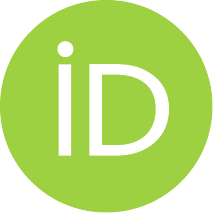}\hspace{1mm}Amandla Mabona$^*$} \\
	InstaDeep Ltd\\
	5 Merchant Square\\
	London, W2 1AY, UK \\
    \texttt{a.mabona@instadeep.com} \\
	\And
	\href{https://orcid.org/0000-0002-1964-3976}{\includegraphics[scale=0.06]{orcid.pdf}\hspace{1mm}Jemma Daniel$^*$} \\
	InstaDeep Ltd\\
	5 Merchant Square\\
	London, W2 1AY, UK \\
    \texttt{j.daniel@instadeep.com} \\
        \And
	\href{https://orcid.org/0009-0006-6039-7339}{\includegraphics[scale=0.06]{orcid.pdf}\hspace{1mm}Henrik Servais Janssen Knudsen} \\
	Department of Biotechnology and Biomedicine\\
	Technical University of Denmark\\
	Kgs. Lyngby, 2100, Denmark \\
    \texttt{s215065@student.dtu.dk} \\
    \And
	\href{https://orcid.org/0000-0002-2983-9833}{\includegraphics[scale=0.06]{orcid.pdf}\hspace{1mm}Rachel Catzel} \\
	InstaDeep Ltd\\
	5 Merchant Square\\
	London, W2 1AY, UK \\
	\texttt{r.catzel@instadeep.com} \\
    \And
	\href{https://orcid.org/0000-0003-1355-8743}{\includegraphics[scale=0.06]{orcid.pdf}\hspace{1mm}Kevin Michael Eloff} \\
	InstaDeep Ltd\\
	5 Merchant Square\\
	London, W2 1AY, UK \\
    \texttt{k.eloff@instadeep.com} \\
    \And
	\href{https://orcid.org/0000-0002-3117-7832}{\includegraphics[scale=0.06]{orcid.pdf}\hspace{1mm}Erwin M. Schoof} \\
	Department of Biotechnology and Biomedicine\\
	Technical University of Denmark\\
	Kgs. Lyngby, 2100, Denmark \\
    \texttt{erws@dtu.dk} \\
    \And
	\href{https://orcid.org/0000-0003-2235-5753}{\includegraphics[scale=0.06]{orcid.pdf}\hspace{1mm}Nicolas Lopez Carranza} \\
	InstaDeep Ltd\\
	5 Merchant Square\\
	London, W2 1AY, UK \\
    \texttt{n.lopezcarranza@instadeep.com} \\
	\And
	\href{https://orcid.org/0000-0003-2979-5663}{\includegraphics[scale=0.06]{orcid.pdf}\hspace{1mm}Timothy P. Jenkins} \\
	Department of Biotechnology and Biomedicine\\
	Technical University of Denmark\\
	Kgs. Lyngby, 2100, Denmark \\
	\texttt{tpaje@dtu.dk} \\
    \And
	\href{https://orcid.org/0000-0003-4480-5567}{\includegraphics[scale=0.06]{orcid.pdf}\hspace{1mm}Jeroen Van Goey$^+$} \\
	InstaDeep Ltd\\
	5 Merchant Square\\
	London, W2 1AY, UK \\
    \texttt{j.vangoey@instadeep.com} \\
    \And
	\href{https://orcid.org/0000-0003-3907-9281}{\includegraphics[scale=0.06]{orcid.pdf}\hspace{1mm}Konstantinos Kalogeropoulos$^+$} \\
	Department of Biotechnology and Biomedicine\\
	Technical University of Denmark\\
	Kgs. Lyngby, 2100, Denmark \\
	\texttt{konka@dtu.dk} \\
}

\date{}

\renewcommand{\headeright}{Preprint}
\renewcommand{\undertitle}{Preprint}
\renewcommand{\shorttitle}{\textit{De novo} peptide sequencing rescoring and FDR estimation with Winnow}

\hypersetup{
pdftitle={Winnow: Post-processing in \textit{de novo} peptide sequencing},
pdfsubject={Mass spectrometry, de novo peptide sequencing},
pdfauthor={Amandla Mabona, Jemma Daniel, Konstantinos Kalogeropoulos},
pdfkeywords= {{\textit{de novo} peptide sequencing}, {mass spectrometry}, {proteomics},
{InstaNovo}, {false discovery rate}, {deep learning}}}

\begin{document}
\maketitle

\begin{center}
\small
* These authors contributed equally to this study.\\[0.5\baselineskip]
+ To whom correspondence should be addressed.\\[2.0\baselineskip]
\end{center}

\begin{abstract}

Machine learning has markedly advanced \textit{de novo} peptide sequencing (DNS) for mass spectrometry-based proteomics.
DNS tools offer a reliable way to identify peptides without relying on reference databases, extending proteomic analysis and unlocking applications into less-charted regions of the proteome.
However, they still face a key limitation. 
DNS tools lack principled methods for estimating false discovery rates (FDR) and instead rely on model-specific confidence scores that are often miscalibrated. 
This limits trust in results, hinders cross-model comparisons and reduces validation success.
Here we present Winnow, a model-agnostic framework for estimating FDR from calibrated DNS outputs.
Winnow maps raw model scores to calibrated confidences using a neural network trained on peptide-spectrum match (PSM)-derived features.
From these calibrated scores, Winnow computes PSM–specific error metrics and an experiment-wide FDR estimate using a novel decoy-free FDR estimator.
It supports both zero-shot and dataset-specific calibration, enabling flexible application via direct inference, fine-tuning, or training a custom model.
We demonstrate that, when applied to InstaNovo predictions, Winnow's calibrator improves recall at fixed FDR thresholds, and its FDR estimator tracks true error rates when benchmarked against reference proteomes and database search.
Winnow ensures accurate FDR control across datasets, helping unlock the full potential of DNS.

\end{abstract}


\section{Introduction}

Bottom-up proteomics has transformed biological research, enabling large-scale proteome analysis through peptide mass spectrum (PSM) identification \cite{aebersold_mass-spectrometric_2016}.
There are two main approaches to PSM identification: database search and \textit{de novo} sequencing (DNS).
Once candidate peptides are assigned, false discovery rate estimation (FDR) is used to retain reliable identifications.

Database search involves matching experimental spectra to theoretical spectra derived from candidate peptides, then using decoy sequences to estimate FDR \cite{sadygov_large-scale_2004, elias_target-decoy_2010, chick_mass-tolerant_2015}.
While effective, this framework faces growing challenges as experimental scale and proteome complexity increase. 
As databases expand to include multiple proteomes and diverse post-translational modifications (PTMs), FDR inflation becomes a significant concern, as the enlarged search space increases random matches and artificially elevates false discovery estimates. \cite{savitski_scalable_2015, freestone_re-investigating_2023}.
At the same time, FDR is frequently underestimated, resulting in overly optimistic thresholds \cite{wen_assessment_2025}. 
Consequently, decoy-free approaches (DFAs) have been proposed to address these issues by modelling correct and incorrect matches as separate distributions and using mixture models to estimate error rates \cite{keller_empirical_2002, gonnelli_decoy-free_2015, peng_algorithm_2024, peng_new_2020, madej_modeling_2023, huang_development_2020}.
Despite their promise, widespread DFA adoption has been hindered by implementation complexity, lack of integration into standard proteomics software platforms, and resource demands.
Furthermore, the distributional assumptions underpinning many of these approaches can degrade the fit of the mixture models across diverse datasets.
DFAs can be used in DNS, which cannot benefit from target-decoy approaches, yet DFAs also tend to be overly conservative.
These methods often assume that low-scoring PSMs are predominantly false positives.
However, in DNS workflows score distributions can be less well separated, containing substantial numbers of mid- and low-scoring true positives.
When these true positives are misclassified as false matches, the resulting FDR is overestimated.

DNS provides an alternative to database search by inferring peptide sequences directly from tandem mass spectra, without relying on reference proteomes \cite{ma_peaks_2003, frank_pepnovo_2005}.
This approach is especially valuable for characterising peptides from poorly annotated species, novel protein variants or previously unobserved PTMs.
In such settings, database searches may be infeasible or results thereof incomplete.
Hybrid approaches that combine DNS with partial database search offer a promising compromise by capturing both known and novel peptides \cite{muth_evaluating_2018, muth_potential_2018}.
However, since many DNS predictions still lack reference sequences, these methods inherit DNS's central weakness: the absence of well-established scoring and FDR estimation strategies.
More broadly, proteomics researchers rely on statistical measures like q-values and posterior error probabilities (PEPs), quantities that allow clearer interpretation and control of error rates \cite{kall_posterior_2008}. 
Such measures remain largely absent in DNS methods, limiting their adoption.

Recent progress in DNS has been driven by machine learning, particularly transformer-based models \cite{qiao_computationally_2021, mao_mitigating_2023, yilmaz_sequence--sequence_2024, zhao_transformer-based_2025, zhang_-primenovo_2025, lee_bidirectional_2024, yang_introducing_2024, eloff_instanovo_2025}.
These models can predict peptide sequences and PTMs from spectra with high recall and generate real-valued confidence scores based on token probabilities.
While such scores are useful for ranking predictions, they are often poorly calibrated \cite{tran_novoboard_2024, qiu2025universalbiologicalsequencereranking}.
This disconnect between model confidence and the true probability that a prediction is correct undermines their use in FDR estimation; we cannot trust these scores as accurate confidence levels.
Furthermore, lack of calibration makes it challenging to compare or integrate predictions across different models.
In previous work \cite{eloff_instanovo_2025}, we used FDR estimates grounded in database search results to identify a confidence score threshold for DNS outputs.
However, this approach depends on the availability of a database and cannot be applied to novel spectra lacking reference matches.
Moreover, when applied to the unlabelled spectra, the shift from the labelled to unlabelled domain often leads to overly optimistic FDR estimates.

In this study, we propose Winnow, a general-purpose rescoring, calibration and FDR estimation framework for peptide identification.
Winnow transforms confidence scores into well-calibrated error probabilities using a supervised calibration model.
Our approach incorporates experimental spectrum features and DNS model inference output to improve prediction reliability and provide familiar metrics in proteomics data.
Crucially, Winnow enables lightweight and statistically rigorous FDR estimation without relying on distributional assumptions or reference databases--a valuable addition to deep learning-based DNS workflows.
Furthermore, our reformulation of FDR using a discriminative decomposition represents, to our knowledge, an entirely novel contribution to the proteomics field.
Winnow corrects miscalibrated confidence scores, improves recall and maintains accurate FDR control.
In doing so, Winnow addresses a key limitation in existing DNS workflows, offering an accurate, trustworthy and model-agnostic method for FDR control that improves generalisation across diverse proteomic landscapes.

\section{Results}
\label{sec:headings}

\subsection{Modelling FDR with Winnow}

We set out to create a generally applicable, database- and decoy-free method for estimating FDR in deep learning-based DNS.
Existing DNS models output amino acid probabilities at each sequence position, which can be aggregated into an overall sequence score, using the dot product or mean probability across positions \cite{eloff_instanovo_2025, yilmaz_sequence--sequence_2024}.
Though model-specific, these scores can be interpreted as confidence estimates for a given PSM.

\begin{figure}[htbp]
    \centering
    \includegraphics[width=1\textwidth]{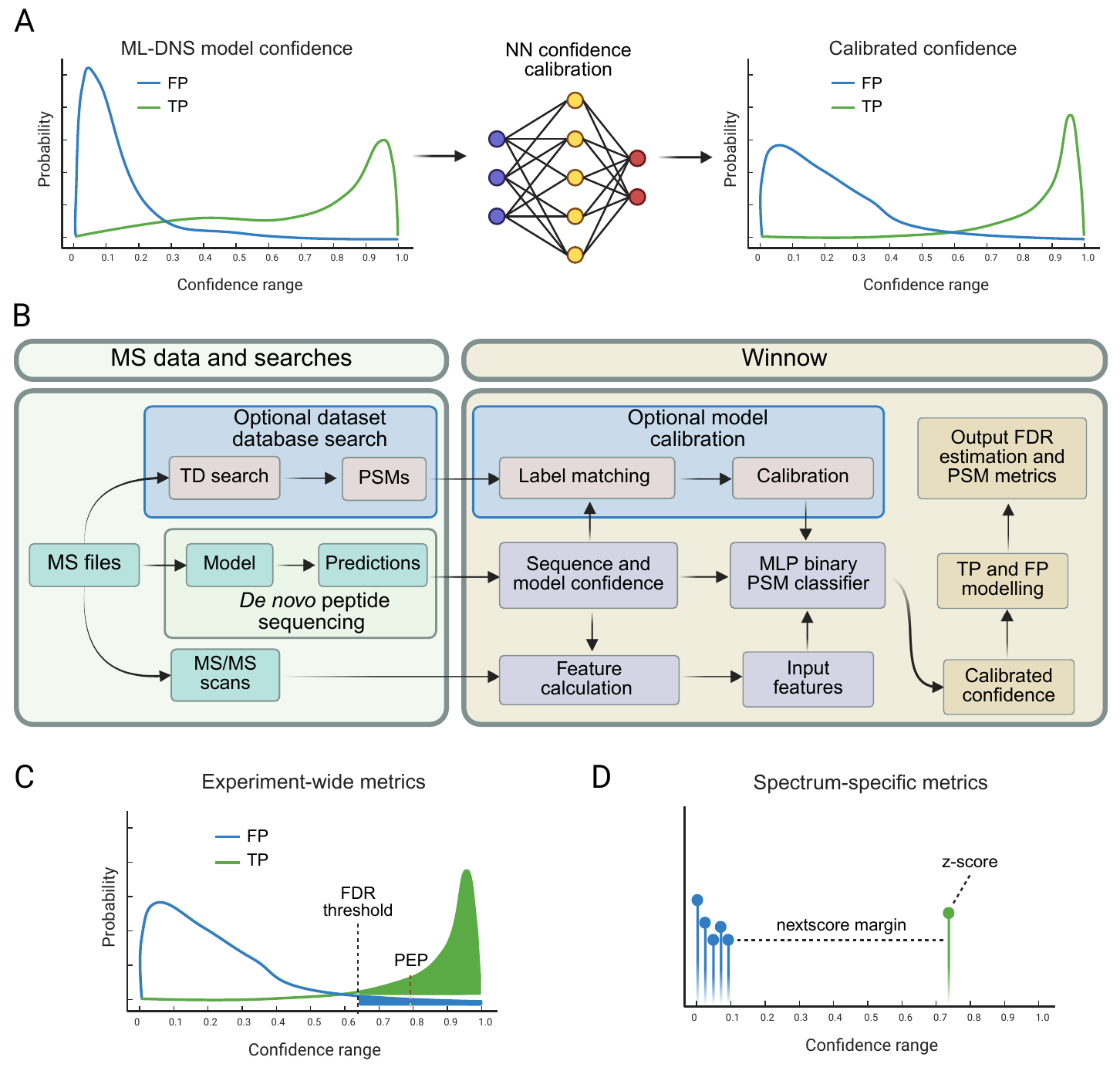}
    \caption{\textbf{Overview of the Winnow framework for FDR estimation in DNS.} \textbf{A}) At the core of the Winnow algorithm is a calibrator model that predicts the likelihood of a PSM being correct, based on features derived from both model outputs and experimental spectra. The model used database search labels for training. Score calibration allows us to estimate FDR and other metrics more accurately, retrieve more correct predictions at lower FDR, and generalise the scoring strategy across models and datasets. \textbf{B}) Schematic showing standard usage of Winnow. The tool takes MS/MS scan information (precursor mass, precursor charge, mass-to-charge and intensity values), DNS predictions, and optionally a database search result for the same MS files for calibration. The Winnow framework includes feature calculation, database label matching and calibration, if selected. Winnow then applies a neural network to assign probabilities of each PSM being correct. This calibrated confidence is then used to non-parametrically estimate FDR. \textbf{C}) The experiment-wide error metrics, FDR and PEP, and \textbf{D}) spectrum-specific metrics are calculated and reported by Winnow, allowing filtering at both levels. Figure made with Biorender.com.}
    \label{fig:winnow_overview}
\end{figure}

Prior decoy-free FDR estimation methods have relied on fitting separate score distributions for correct and incorrect identifications, which requires both a choice of distributional form and a validation of fit.
Instead, we took a discriminative approach: directly learning the probability that a given PSM is correct using a calibrated binary classifier.
We used this solution to create Winnow, a calibrate-then-estimate method for FDR control in deep learning-based DNS workflows.
Winnow trains a calibration model on database search results, learning to map DNS-model-generated confidence scores, alongside additional mass spectrometry features, to calibrated probabilities of correctness (Fig. \ref{fig:winnow_overview}A).
These calibrated scores serve as inputs to our FDR estimator which relies only on calibrated confidence outputs, sidestepping the need for fitted PSM correctness distributions.

The Winnow pipeline consists of four key stages (Fig. \ref{fig:winnow_overview}B).
\begin{enumerate}
    \item \textbf{Input processing and sequencing}
    \begin{description}
        \item[] Raw mass spectrometry data are processed by a sequencing or search model that generates PSMs and initial confidence scores.
    \end{description}
    \item \textbf{Feature computation}
    \begin{description}
        \item[] A set of supplementary features are computed for each prediction, including precursor mass errors, the number and intensity of matches between predicted fragment ions and observed spectra, retention time error and, where available for deep learning-based DNS models, beam search statistics.
    \end{description}
    \item \textbf{Calibration}
    \begin{description}
        \item[] The raw confidence scores are recalibrated using a neural network classifier that learns to map the computed features and raw confidences to well-calibrated probabilities.
    \end{description}
    \item \textbf{FDR estimation}
    \begin{description}
        \item[] The calibrated probabilities are used to estimate FDR using one of two approaches. In the first, a label-free, non-parametric method estimates error rates directly from the calibrated confidence scores without assuming a specific distribution. In the second, database search results are used to distinguish likely correct from likely incorrect model identifications.
    \end{description}
\end{enumerate}

In addition to FDR estimation, Winnow also supports the computation of familiar metrics in proteomics, such as PEP, q-value and nextscore (Fig. \ref{fig:winnow_overview}C; Fig. \ref{fig:winnow_overview}D).
In doing so, we move beyond raw confidence scores to provide a more trustworthy method of FDR control in DNS settings.

We implemented Winnow as a flexible, extensible Python package that supports customisable feature selection for calibration, builds on scikit-learn for calibration models, and makes available our lightweight, non-parametric approach to FDR estimation.

We additionally enable zero-shot FDR estimation by providing a general calibration model trained on InstaNovo predictions from a wide variety of spectral datasets (Supplementary Fig. \ref{fig:dataset_characteristics}A).

\subsection{Selecting and engineering features for optimal model calibration} \label{sec:feature_selection}

\begin{figure}[htbp]
    \centering
    \includegraphics[width=1\linewidth]{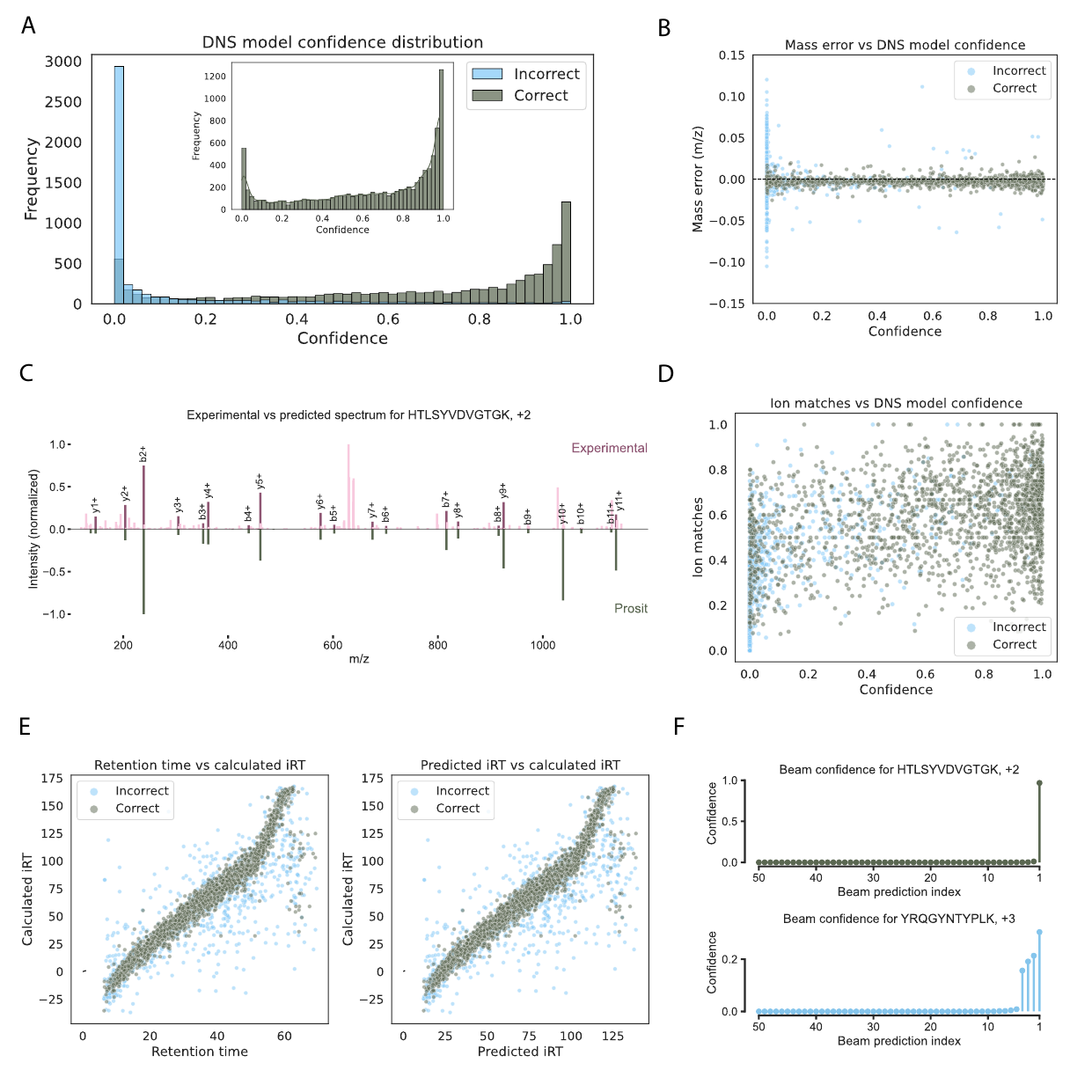}
    \caption{\textbf{Feature selection for accurate correct PSM determination.} \textbf{A}) PSM confidence distribution in a standard DNS experiment. Confidence is the dot product of the exponentiated amino acid-level log probabilities for the peptide sequence predicted by the model. Correct PSMs are clustered in the high confidence range, while predictions for spectra not containing peptides are assigned low confidence. All plots show results from the HeLa Single Shot dataset test set and are labelled by PSM correctness according to database search results. \textbf{B}) Relationship between confidence and mass error. Low confidence predictions exhibit higher mass errors. \textbf{C}) Prosit predicted fragment ion profiles are used to compute percent ion intensity present in the experimental spectrum for the predicted sequence. \textbf{D}) Relationship between normalised number of ion matches of predicted sequences with PSM confidence. \textbf{E}) Relationship between calculated iRT with experimental retention times (left) and Prosit predicted iRT (right). \textbf{F}) Example beam confidence value distributions for a correct prediction with high confidence (top) and an incorrect prediction with low confidence (bottom) PSM.}
    \label{fig:hela_feature_selection}
\end{figure}

While DNS model confidence scores show good discrimination between correct and incorrect PSMs, as judged by database labels, many correct predictions still occur across the full confidence range (Fig. \ref{fig:hela_feature_selection}A). 
At the same time, FDR remains high even at the upper end of the confidence scale, due to a subset of incorrect PSMs that are assigned high model confidence.
This limits the ability to recover true positives using a simple confidence threshold and motivates the need for better calibrated confidence estimates that integrate additional information.

Previous work in database search and DNS has shown that rescoring PSMs with additional mass spectrometry (MS)-derived features, beyond raw model or search engine score, can substantially improve the number of identifications recovered at a given FDR \cite{yang_msbooster_2023, kalhor_rescoring_2024, miller_postnovo_2018}.
Consequently, we selected a set of features that could be used, adapted or computed for the DNS setting.

First, we considered features that quantify agreement between predicted sequences and experimental observations.
These include precursor mass error, the number of matched fragment ions and the percentage of spectrum intensity explained by matched peaks.
Precursor mass error gives the difference between the experimental and theoretical precursor mass and tends to be larger for incorrect predictions. Indeed, we observed that extremely low model confidences are often associated with large mass error values (Fig. \ref{fig:hela_feature_selection}B).
We additionally computed fragment ion match rate and intensity by comparing observed spectra against Prosit-predicted spectra from the DNS peptide identification (Fig. \ref{fig:hela_feature_selection}C).
Incorrect sequences typically resulted in fewer ion matches and lower explained intensity (Fig. \ref{fig:hela_feature_selection}D).
To capture retention time agreement, we learned a mapping between retention time and iRT across experiments and computed this difference for each prediction.
We found good correlation between predicted and estimated iRT values for correct identifications and increased variance between these values for incorrect identifications (Fig. \ref{fig:hela_feature_selection}E).

We also included features derived from the beam of candidate sequences predicted for each spectrum.
These include the margin between the confidence score of the best and second-best prediction (also known as the nextscore), and a z-score representing the top prediction's confidence relative to the distribution across beam candidates (Fig. \ref{fig:hela_feature_selection}F).

Together, these features are useful discriminators for better separating correct and incorrect PSMs, providing additional resolution when model confidence alone is ambiguous.
These features, alongside model confidence, form the input space for our calibration model, which sits at the core of Winnow.

\subsection{Training a model for PSM score calibration}

To estimate FDR in deep learning-based DNS, we previously introduced a method we refer to as \textit{database-grounding}, which uses database searches as a surrogate for the ground truth peptide identification \cite{eloff_instanovo_2025}.
While these labels are not perfect, they represent the most reliable approximation of DNS prediction correctness currently available.
Building on this idea, we utilised the available sequence labels afforded by database searches to train a binary classifier to distinguish between correct and incorrect PSMs predicted by DNS.
Our confidence calibration model is trained on a variety of features derived from both mass spectrum and DNS inference outputs, learning to predict the likelihood that a given PSM is a correct assignment.

\begin{figure}[htbp]
    \centering
    \includegraphics[width=0.9\linewidth]{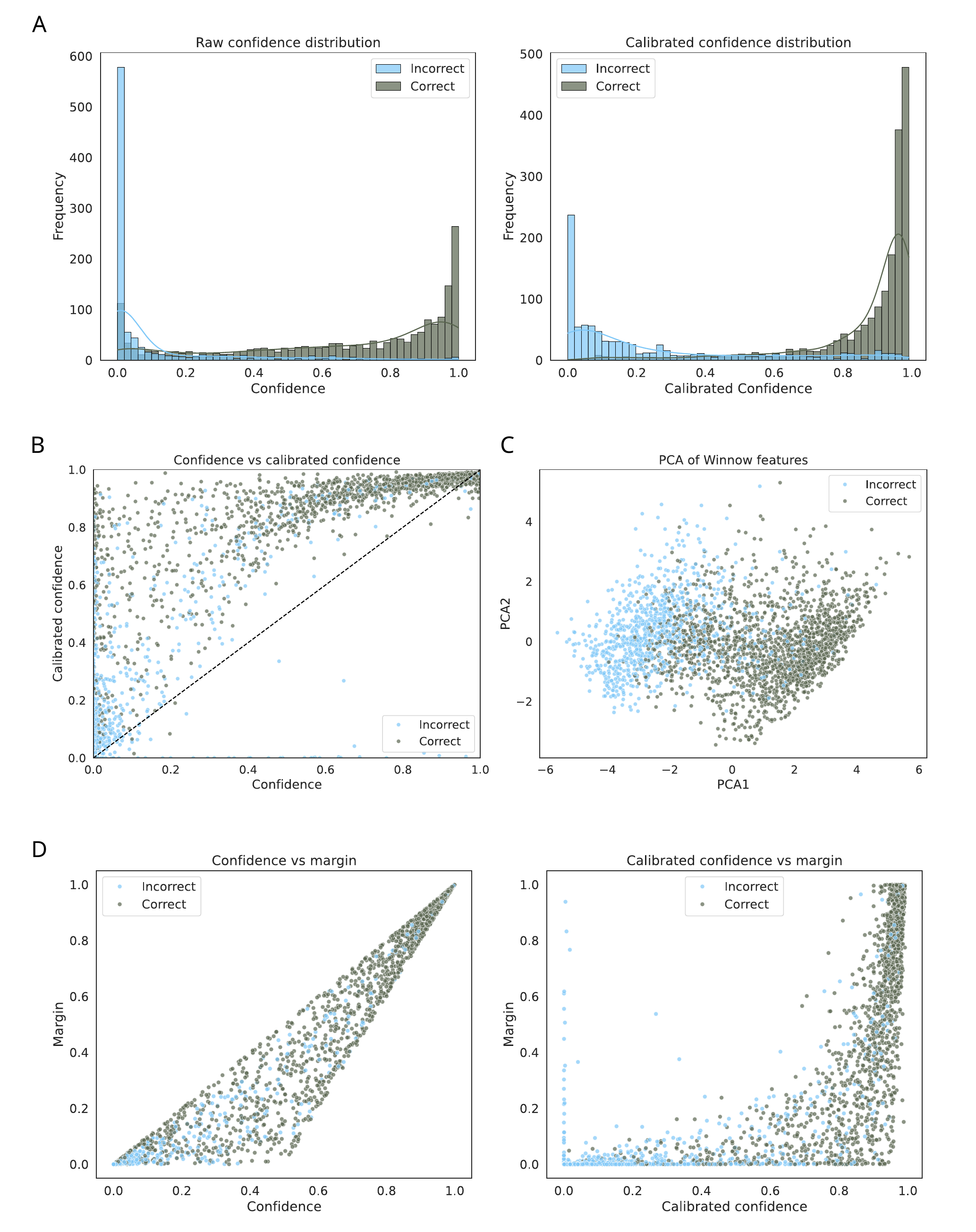}
    \caption{\textbf{Winnow PSM rescoring in DNS for HeLa Single Shot proteome dataset.} \textbf{A}) Rescoring and calibration of DNS peptide predictions across the confidence range, labelled for correctness according to database search results. The output confidence values (left) are rescored with Winnow to produce calibrated confidence values (right), grounded by MS properties and database search results. \textbf{B}) Mapping of confidence value shift before and after calibration. \textbf{C}) Visualisation of the first two principal components for the features used in Winnow training and inference. \textbf{D}) Relationship between the margin, the confidence difference between best and second best sequence prediction in the beam, and PSM confidence, before calibration (left) and after calibration (right).}
    \label{fig:hela_confidence_calibration}
\end{figure}

We evaluated our approach using a model trained on the labelled HeLa Single Shot dataset for the spectra that received database matches.
The raw confidence scores from the DNS model were skewed towards underconfidence, with a large spike in the low confidence region (Fig. \ref{fig:hela_confidence_calibration}A). 
After calibration with Winnow, the distribution became more separable and better aligned with empirical accuracy.
Our calibrator shifted a substantial portion of the predictions to higher score regions, indicating that InstaNovo could be overly conservative when compared to the true probability of PSM correctness (Fig. \ref{fig:hela_confidence_calibration}B).
A comparison of internal representations after calibration supports this improvement: correct and incorrect PSMs were largely separable in PCA space following calibration, which indicates that the calibrated scores faithfully captured underlying relationships relevant to PSM correctness (Fig. \ref{fig:hela_confidence_calibration}C).
Importantly, the resulting confidence score generalises to unlabelled DNS predictions, unlocking calibrated confidence estimation even in regions not covered by database matches.

Exploratory PCA revealed that PC1 was primarily driven by raw DNS model confidence, median margin and margin, while PC2 was largely driven by ion match intensity and chimeric ion match intensity, indicating that the main axes of variance separate correct PSM identifications by model-derived confidence scores and orthogonally by spectral matching evidence (Supplementary Fig. \ref{fig:dataset_characteristics}B).

We also examined how confidence relates to PSM prediction margin--the difference between the top-1 and top-2 beam candidates (Fig. \ref{fig:hela_confidence_calibration}D).
Raw confidence possessed a strong positive correlation with margin.
All values are contained within a sharply delineated region due to the beam prediction constraint in InstaNovo, which requires the scores on a beam to sum to one.
The calibrator tempers the influence of margin; correlation between margin and calibrated confidence remained present but not as strong.
At high margins, calibrated confidence still approached 1.0, as expected--indicating that the calibrator assigned high scores for PSMs where the DNS model was strongly confident.
However, the calibrator shifted some spectra with correct PSM identifications yet low margins to higher confidence ranges.

Counterintuitively, higher chimeric ion match intensity occurred more frequently in correct PSMs than in incorrect ones. 
In the raw confidence space, chimeric ion match intensity was uncorrelated with DNS model confidence and did not separate correct from incorrect identifications (Supplementary Fig. \ref{fig:chimeric}A).
After calibration, correct and incorrect PSMs are separated along the calibrated probability axis.
Correct PSMs clustered at high confidence with low-to-moderate intensity values, while incorrect PSMs clustered in the low-confidence region with low intensity (Supplementary Fig. \ref{fig:chimeric}B). 
Kernel density estimates confirm this pattern: incorrect PSMs mostly have low intensity, with only a small tail of incorrect spectra reaching moderate intensities (possibly indicating the presence of a few chimeric spectra), whereas correct PSMs are broadly distributed across moderate intensities (Supplementary Fig. \ref{fig:chimeric}C).
This indicates that the feature's signal is context-dependent and requires supporting signal from other features, particularly at low values.
Altogether, we observe that Winnow's calibration model improved the discriminative and probabilistic quality of the DNS model's PSM predictions.

\subsection{Winnow estimates FDR in DNS settings without prior assumptions}

We evaluated Winnow's FDR estimation on the labelled subset of the HeLa Single Shot proteome dataset.
To fairly position Winnow against existing practice, we compared against two database-grounded baselines: database-grounded FDR on raw DNS confidences, and database-grounded FDR on calibrated confidences (to isolate the effect of the estimation method itself).
Winnow's non-parametric, label-free procedure was applied only to calibrated confidences, since accurate FDR estimation in this framework requires well-calibrated scores.

\begin{figure}[htbp]
    \centering
    \includegraphics[width=1\textwidth]{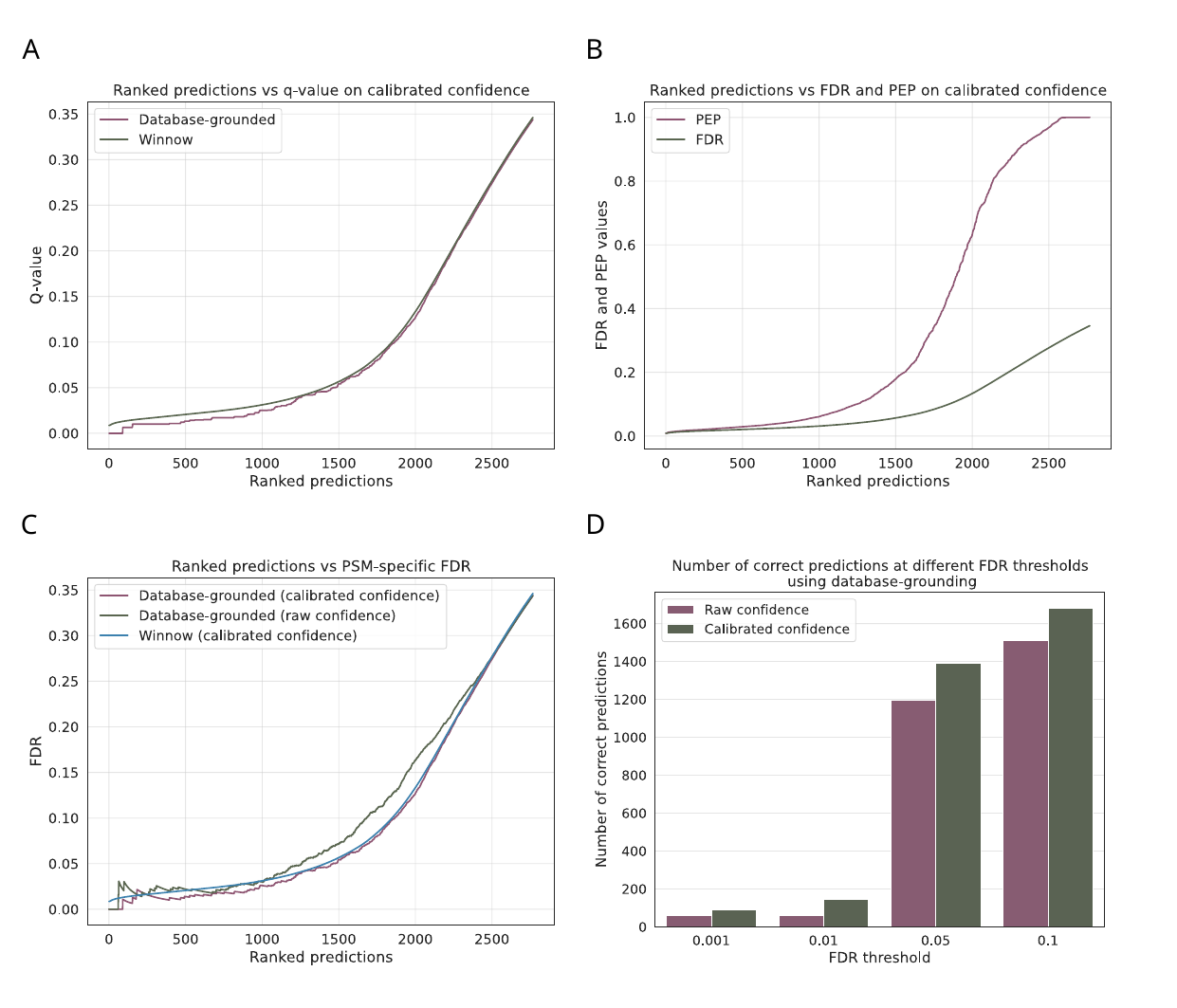}
    \caption{\textbf{Performance of Winnow's FDR estimation method on the HeLa Single Shot dataset.} \textbf{A}) Comparison between q-values produced by Winnow and q-values computed by database-grounding. Predictions are ranked in descending order of calibrated confidence. \textbf{B}) Profile of FDR and PEP values in ranked predictions. \textbf{C}) Profiles of FDR when using Winnow, using database-grounding with calibrated confidence, and using database-grounded with raw DNS model confidence. \textbf{D}) Non-parametric FDR estimation on calibrated confidence retrieves more predictions at different FDR thresholds compared to database-grounded FDR estimation with raw DNS model confidence.}
    \label{fig:hela_fdr_performance}
\end{figure}

Winnow supports estimation of both PEP and q-values, the PSM-specific error metrics commonly used in peptide identification pipelines.
Q-values obtained from our non-parametric method closely tracked those computed with ground-truth labels (Fig. \ref{fig:hela_fdr_performance}A), demonstrating the reliability of our label-free estimation.
Because our approach enforces monotonicity, PSM-specific FDR values are by construction equivalent to q-values.

We next compared PEP and PSM-specific FDR for ranked predictions using Winnow's non-parametric method (Fig. \ref{fig:hela_fdr_performance}B).
Here, PEP is the complement of calibrated confidence and spans the full 0–1 range.
As expected, PEP rose more steeply than FDR, reflecting its interpretation as a local error probability.
In contrast, FDR accumulates errors across sets of PSMs, making it consistently more permissive than PEP for individual identifications \cite{kall_posterior_2008}.

We then proceeded to compare PSM-specific FDR estimates from database-grounded and non-parametric methods (Fig. \ref{fig:hela_fdr_performance}C).
Database-grounded FDR is calculated independently at each score threshold and is therefore not monotonic; local fluctuations in true and false positives cause oscillations in the curve.
Winnow's non-parametric estimates, in contrast, are monotonic and closely overlapped with database-grounded results on calibrated scores, showing highly accurate FDR control.
Using database-grounded FDR on raw DNS model confidence results in larger FDR estimates for mid-ranked PSMs, which indicates that Winnow's calibrator better separates ambiguous cases where the DNS model is uncertain.

Finally, we assessed the number of correct PSMs identified at standard FDR thresholds across the two approaches.
Winnow's full calibrate-then-estimate pipeline consistently recovered more correct identifications (Fig. \ref{fig:hela_fdr_performance}D; Supplementary Table \ref{tab:hela_results}). 
In summary, Winnow's calibration and non-parametric FDR estimation provided reliable FDR control while increasing recall, yielding more PSMs at commonly used FDR thresholds.

\subsection{Robust FDR estimation assigns statistical confidence to DNS predictions}

\begin{figure}[htbp]
    \centering
    \includegraphics[width=0.9\linewidth]{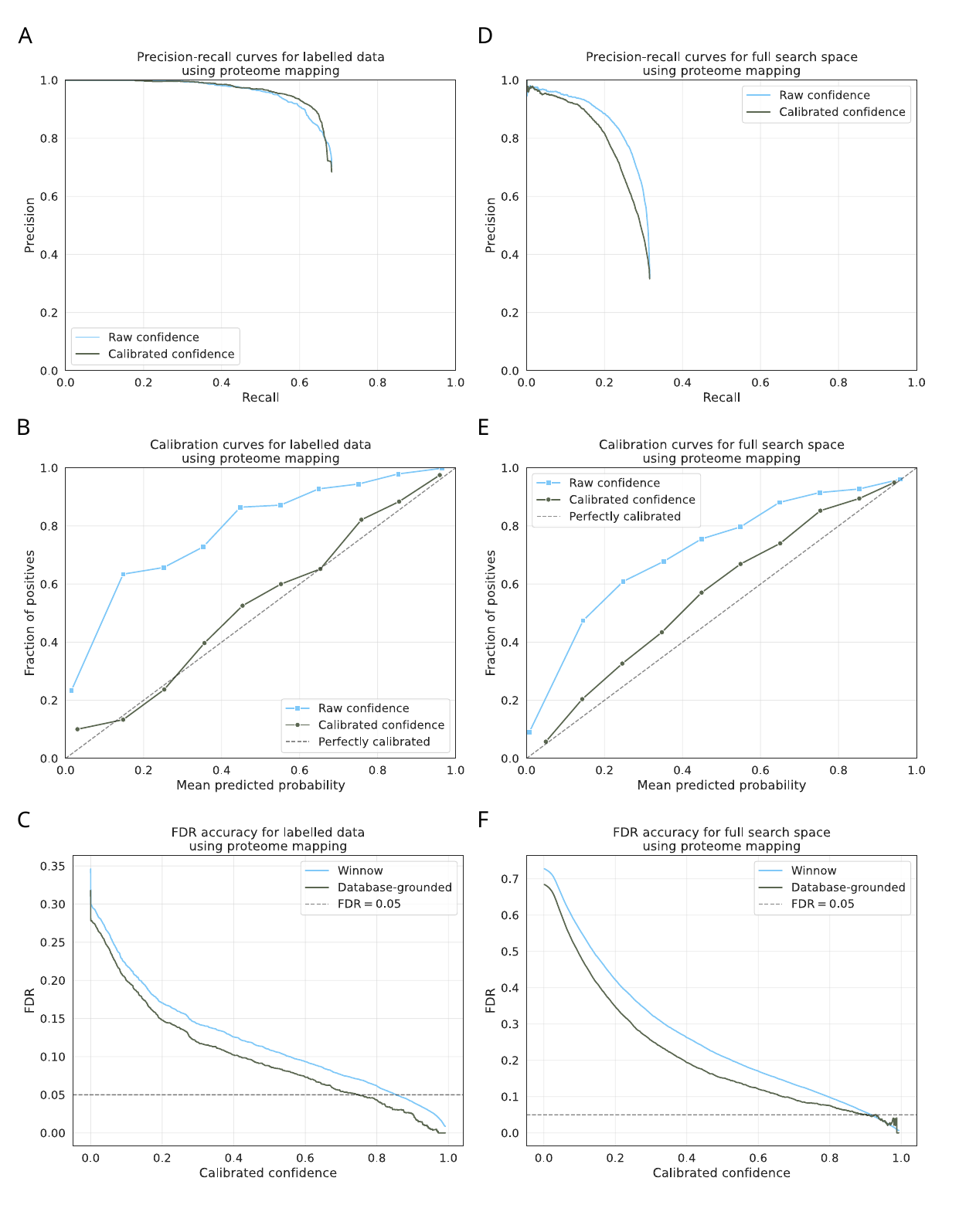}
    \caption{\textbf{Performance of Winnow's full pipeline on the HeLa Single Shot dataset using proteome hits as a proxy for PSM correctness via database search.} \textbf{A}) Precision-recall curves comparing calibrated and raw DNS model confidence on the labelled test set of HeLa Single Shot. \textbf{B}) Calibration curves for calibrated and raw confidence, compared against perfect calibration, for the labelled test set. \textbf{C}) PSM-specific FDR run plots for Winnow's non-parametric and database-grounded FDR estimation methods, using calibrated confidence, for the labelled test set. \textbf{D}) Precision-recall curves for calibrated and raw confidence on the full HeLa Single Shot dataset less the training set. \textbf{E}) Calibration curves for calibrated and raw confidence over the full HeLa dataset less the training set. \textbf{F}) PSM-specific FDR run plots for the full HeLa dataset less the training set using non-parametric FDR estimation and database-grounded FDR estimation on calibrated DNS model confidence.}
    \label{fig:hela_proteome_mapping_performance}
\end{figure}

DNS predictions that do not match a reference database cannot be directly evaluated for correctness, making accurate FDR estimation in the unlabelled space particularly challenging.
To benchmark our proposed calibrate-then-estimate approach in the absence of ground truth, we used human proteome hits as a proxy for correctness on the HeLa Single Shot dataset as previously established \cite{eloff_instanovo_2025, sanders_transformer_2024}.
While not definitive, this surrogate enabled us to empirically assess model calibration and FDR control performance.
This approximation is appropriate in this context because the HeLa cell line is well-characterised and the human proteome has been annotated comprehensively.
We assessed this by plotting precision-recall, calibration and FDR accuracy for the labelled test set of HeLa Single Shot using correct proteome hit in place of PSM correctness via database search.
Comparable performance between the use of proteome mapping- and database search-based labels for the labelled subset confirmed that proteome mapping is a reasonable label proxy.
The precision-recall curves showed that calibrated confidence yielded higher recall at equivalent precision levels compared to raw DNS model confidence (Fig. \ref{fig:hela_proteome_mapping_performance}A; Supplementary Fig. \ref{fig:hela_test_with_database_search}A). 
Winnow's calibrated confidences proved better aligned with empirical probabilities than raw confidences (Fig. \ref{fig:hela_proteome_mapping_performance}B; Supplementary Fig. \ref{fig:hela_test_with_database_search}B).
Additionally, the FDR run plot for Winnow's non-parametric method closely tracked FDR estimation using database-grounding when both computed on calibrated confidence to isolate the behaviour of each FDR method. 
Non-parametric FDR estimation provided stricter PSM-specific FDR estimates than database-grounding when using correct proteome hits as a proxy for correct PSM identification (Fig. \ref{fig:hela_proteome_mapping_performance}C; Supplementary Fig. \ref{fig:hela_test_with_database_search}C).
This discrepancy arises because database-grounding with proteome mapping underestimates the true FDR (calculated using database PSM correctness)--a problem which Winnow circumvents.

Next, we evaluated the Winnow pipeline on the full HeLa Single Shot dataset (excluding the calibrator's training set) to simulate a realistic DNS setting.
Although raw confidence outperformed calibrated confidence for PSM rescoring, Winnow's calibration step has dual purposes: both improving PSM ranking and aligning predicted scores with correctness probabilities to accurately control FDR (Fig. \ref{fig:hela_proteome_mapping_performance}D).
This trade-off is further contextualised by the calibration curves, which show that calibrated scores were significantly closer to perfect calibration than raw scores, supporting more meaningful probabilistic interpretation and more accurate FDR modelling (Fig. \ref{fig:hela_proteome_mapping_performance}E).
Winnow's FDR estimates tracked the database-grounded FDR curve closely and conservatively across all confidence levels, offering reliable error control that avoids overconfident inclusion of false positives (Fig. \ref{fig:hela_proteome_mapping_performance}F).
Poor calibrator ranking performance in this unlabelled space is likely a result of significant label and feature distribution shift between the labelled and unlabelled subsets of the HeLa Single Shot dataset, which could be improved by training on a broader range of datasets to identify globally applicable causal relationships.

In addition, we compare FDR control and recall between the two FDR estimation methods at 5\% FDR (Supplementary Table \ref{tab:hela_results}).
We assessed our previous method of FDR control in DNS settings (i.e., estimating a confidence cutoff on the subset of database-labelled predictions for a dataset, then extrapolating to the full set) against Winnow's novel end-to-end procedure that calibrates DNS confidence and estimates FDR across all predictions, using correct proteome hits as a proxy for true PSM correctness.
Our calibrate-then-estimate approach achieved empirical FDR near the 5\% target, while extrapolating a database-grounded confidence cutoff led to an observed FDR above the target.

In summary, we showed that Winnow provides well-calibrated probabilities, which our label-free, non-parametric FDR estimation procedure used to successfully track FDR estimates obtained via database-grounding, unlocking reliable FDR control in DNS settings.

\subsection{Winnow's pre-trained calibration model generalises to unseen datasets}

Demonstrating that calibration trained on one dataset can generalise to others would indicate that Winnow captures broadly applicable patterns in PSM predictions and experimental metadata.
To address this, we trained a general-purpose calibration model across eight datasets to capture global patterns and enable reliable zero-shot calibration on external DNS predictions.
Evaluation on a labelled test set made up of held-out PSMs from the combined datasets showed increased recall at all precision levels (Supplementary Fig. \ref{fig:general_test_performance}A), near-perfect calibration (Supplementary Fig. \ref{fig:general_test_performance}B), and extremely accurate FDR control (Supplementary Fig. \ref{fig:general_test_performance}C).

\begin{figure}[htbp]
    \centering
    \includegraphics[width=0.8\linewidth]{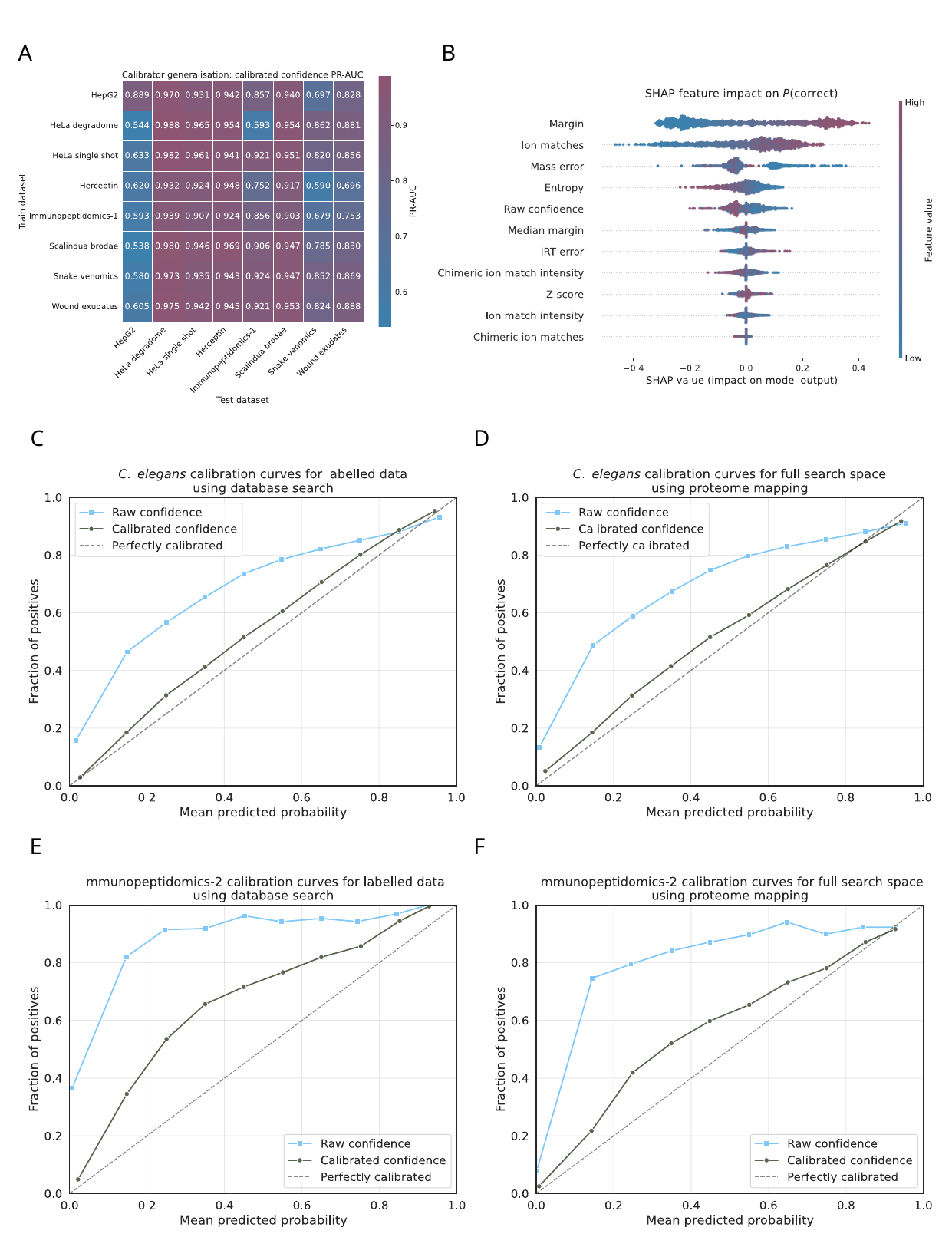}
    \caption{\textbf{Performance of Winnow's pre-trained (or general) calibrator on two held-out datasets: \textit{C. elegans} and Immunopeptidomics-2.} \textbf{A}) Hold-one-out evaluation over all datasets included in the Winnow general model training set. A model is trained on a single given dataset, then evaluated on each of the remaining datasets using area under the precision-recall curve (PR-AUC). \textbf{B}) SHAP beeswarm plots for all features used to train Winnow's general model, ranked by SHAP feature importance. Jittered points show data density over the SHAP value range. \textbf{C}) Calibration curves for the labelled subset of the \textit{C. elegans} dataset, comparing calibrated and raw DNS model confidence. \textbf{D}) Calibration curves for the full \textit{C. elegans} dataset, comparing calibrated and raw DNS model confidence. \textbf{E}) Calibration curves for the labelled subset of the Immunopeptidomics-2 dataset, comparing calibrated and raw DNS model confidence. \textbf{F}) Calibration curves for the full Immunopeptidomics-2 dataset, comparing calibrated and raw DNS model confidence.}
    \label{fig:general_calibrator_perf_and_features}
\end{figure}

To assess generalisation, we also performed a hold-one-dataset-out evaluation across the general calibration model training set.
We trained a calibrator on single dataset and evaluated it zero-shot on the remaining seven held-out datasets, measuring area under the precision-recall curve (PR-AUC) for distinguishing correct from incorrect PSMs (Fig. \ref{fig:general_calibrator_perf_and_features}A).
We also compared this PR-AUC against that achieved by raw DNS model confidence (Supplementary Fig. \ref{fig:extra_feature_importance_plots}A).
Most held-out datasets achieved strong performance (PR-AUC for calibrated confidence between 0.9 and 0.95), indicating robust generalisation across varied contexts.
Performance was lower across all single-dataset models on HepG2, Snake Venomics and Wound Exudates, likely due to systematic differences; HepG2 was acquired on a different generation of instrument, while the Snake Venomics and Wound Exudates data are inherently noisier.
These dataset-specific features limited their generalisability to other contexts, highlighting the need to combine diverse datasets when training a general calibration model.

To better understand our general model's behaviour, we performed feature-level interpretability experiments.
We used SHAP (SHapley Additive exPlanations) to attribute prediction probabilities to input features across 1,000 randomly sampled test set spectra, using a KernelExplainer with 500 background training samples.
We calculated feature importance ranking, feature similarity clustering and feature impact on calibrator output across predictions (Fig. \ref{fig:general_calibrator_perf_and_features}B; Supplementary Fig. \ref{fig:extra_feature_importance_plots}B; Supplementary Fig. \ref{fig:extra_feature_importance_plots}C).
We further computed permutation importance scores by measuring the drop in model performance when each feature was randomly shuffled, averaged over ten runs (Supplementary Fig. \ref{fig:extra_feature_importance_plots}D).
Our explainability experiments indicated that the highest contributing features were margin and ion matches.
These top two features exhibited substantially higher mean absolute SHAP values and importance scores compared to the rest, indicating that they contributed the most to the model's predictions across the dataset (Supplementary Fig. \ref{fig:extra_feature_importance_plots}B; Supplementary Fig. \ref{fig:extra_feature_importance_plots}D).
Echoing our previous findings, we observed that high margins and numbers of ion matches implied a greater probability of PSM correctness (Supplementary Fig. \ref{fig:top_feature_importances}A; Supplementary Fig. \ref{fig:top_feature_importances}B).
Mass error was the third most important feature, however it did not display a strong monotonic relationship between its value and SHAP contribution.
This may be because mass error is not an absolute value, so both high and low values are likely to negatively impact the probability of PSM correctness, and there may be additional non-linear or context-dependent effects influencing calibrator output.

Interestingly, the SHAP profile for the raw DNS model confidence revealed a non-monotonic relationship (Fig. \ref{fig:general_calibrator_perf_and_features}B).
While we may expect higher raw confidence to positively influence the calibrated probability of correctness, the SHAP values show that low raw confidences often received a positive contribution, while high confidences sometimes reduced the final calibrated score.
This was consistent with the overall underconfidence of InstaNovo, our DNS model, and suggests that the calibrator compensated by boosting low-confidence predictions and moderating overly confident ones.

Hierarchical SHAP clustering revealed three groups of features with correlated contributions to model predictions: DNS confidence- and beam-related features, the spectral evidence features ion matches and chimeric intensity, and additional spectral features iRT error and ion match intensity (Supplementary Fig. \ref{fig:extra_feature_importance_plots}B).
This means these clustered features all play similar roles during calibrator prediction.
Furthermore, these features all possessed strong positive pairwise correlations with other features within their respective cluster, implying overlapping raw information and possible redundancy (Supplementary Fig. \ref{fig:top_feature_importances}C).

Although lower-ranked features still displayed consistent and interpretable SHAP patterns, indicating they provide complementary signal.
However, the majority of their values clustered near zero, suggesting the model relied primarily on the top-ranked features, with these features only occasionally influencing predictions (Fig. \ref{fig:hela_feature_selection}E).
Overall, our general model has learned that margin and ion match rate are highly predictive of PSM correctness and that several other features provide similar predictive signals.

We further evaluated generalisation performance on two held-out datasets, \textit{C. elegans} and Immunopeptidomics-2.
We observed near-perfect calibration and thus a strong generalisation ability for the \textit{C. elegans} labelled-only and full datasets (Fig. \ref{fig:general_calibrator_perf_and_features}C; Fig. \ref{fig:general_calibrator_perf_and_features}D).
While still better-aligned with empirical probabilities than raw DNS confidence, calibration was poorer on the Immunopeptidomics-2 dataset (Fig. \ref{fig:general_calibrator_perf_and_features}E; Fig. \ref{fig:general_calibrator_perf_and_features}F).
Precision-recall curves indicated that our general calibrator model increased recall for the same precision when compared to raw DNS model confidence on both hold-out datasets (Supplementary Figs. \ref{fig:external_data_pr_curves}A–D).
In summary, we found improved calibration and generalisability when evaluating our general calibration model on unseen experimental conditions.

\subsection{Winnow controls FDR in external datasets}

\begin{figure}[htbp]
    \centering
    \includegraphics[width=1\linewidth]{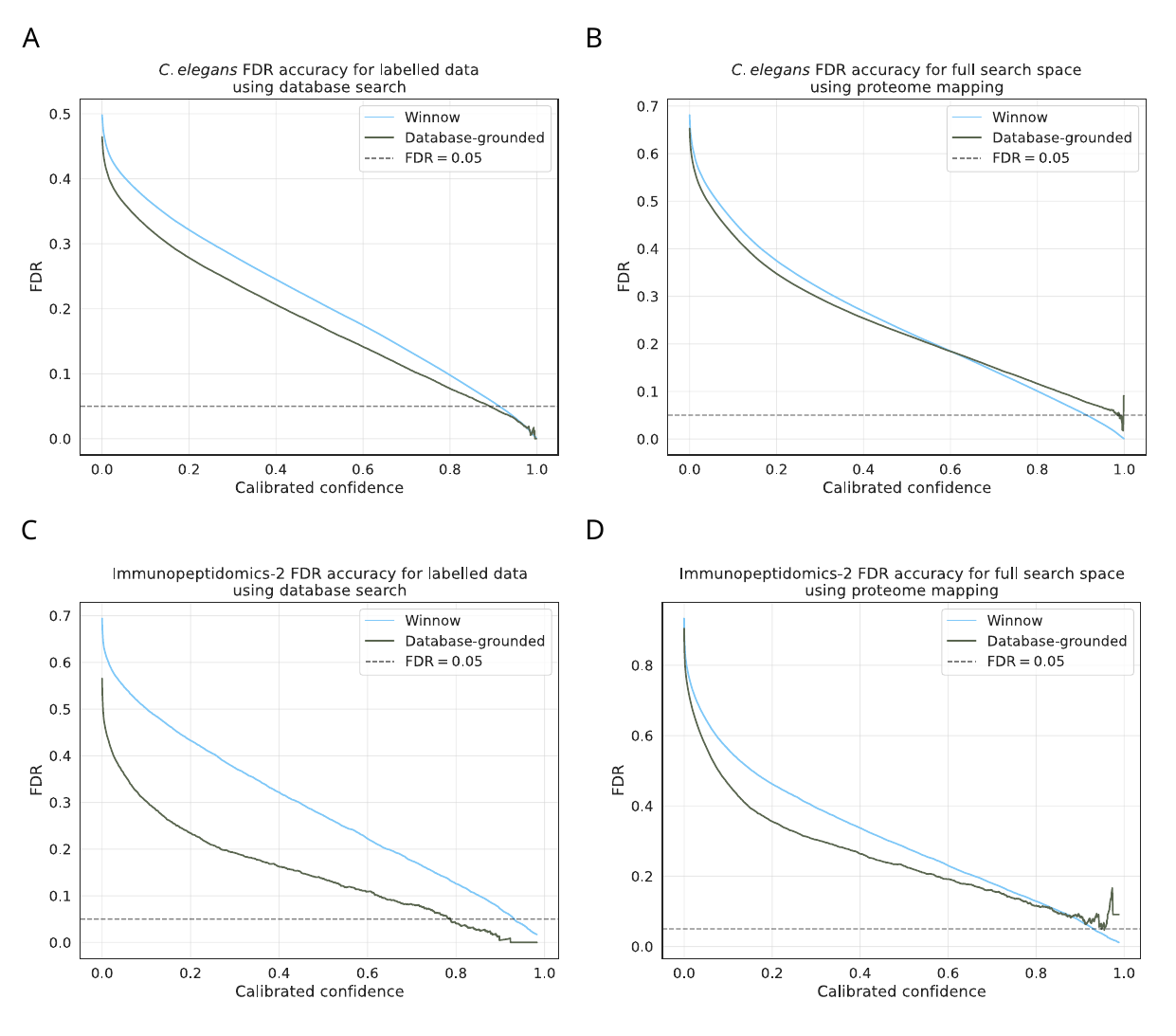}
    \caption{\textbf{Performance of Winnow's FDR estimation methods on two held-out datasets: \textit{C. elegans} and Immunopeptidomics-2.} \textbf{A}) PSM-specific FDR run plots for Winnow's non-parametric and database-grounded FDR estimation methods on the labelled subset of \textit{C. elegans}. \textbf{B}) PSM-specific FDR run plots on the labelled subset of Immunopeptidomics-2, comparing non-parametric and database-grounded FDR estimation. \textbf{C}) PSM-specific FDR run plots for the full \textit{C. elegans} dataset, comparing non-parametric and database-grounded FDR estimation. \textbf{D}) PSM-specific FDR run plots for the non-parametric and database-grounded FDR estimation methods on the full Immunopeptidomics-2 dataset.}
    \label{fig:general_fdr_performance}
\end{figure}

We validated Winnow’s FDR estimation on two held-out datasets, \textit{C. elegans} and Immunopeptidomics-2, that originate from different studies.
This allowed us to assess the model’s robustness to both biological and technical distribution shifts.
For each dataset, we compared Winnow's label-free FDR estimates against estimates derived with database-grounding, using correct proteome hits as a proxy for database PSM correctness for the full, unlabelled, dataset.
We found that our non-parametric FDR estimation method faithfully tracked database-grounded FDR estimation for the \textit{C. elegans} dataset (Fig. \ref{fig:general_fdr_performance}A; Fig. \ref{fig:general_fdr_performance}B).
We further observed that non-parametric FDR estimation yielded conservative results for the Immunopeptidomics-2 dataset in both the labelled and full subsets, caused by poorer calibration (Fig. \ref{fig:general_fdr_performance}C; Fig. \ref{fig:general_fdr_performance}D; Fig. \ref{fig:general_calibrator_perf_and_features}C; Fig. \ref{fig:general_calibrator_perf_and_features}D).

We also recorded empirical recall and FDR at the user-defined 5\% FDR cutoff threshold across Winnow's FDR estimation methods (Supplementary Table \ref{tab:external_results}), using correct proteome hit as a stand-in for correct PSM identification for the full search space and correct PSM in the labelled space.
On the \textit{C. elegans} dataset, Winnow’s pipeline delivered substantially higher recall than the existing approach of applying database-grounded FDR estimation to raw DNS confidences.
Importantly, it did so while maintaining comparable FDR across both labelled and full datasets.
For Immunopeptidomics-2, Winnow yielded more conservative estimates on the labelled set, albeit with lower recall but very tight FDR control. 
Winnow also maintained strong performance in the Immunopeptidomics-2 full search space, where extrapolated database thresholds became overly permissive.

These results highlight Winnow’s ability to deliver reliable FDR control in unseen DNS settings, where label scarcity or biological novelty make extrapolation brittle.

\subsection{Flexible feature design for calibration}

Winnow's calibration system has been designed to be modular and extensible, enabling rapid prototyping and integration of new features without major changes to the core system.
Central to Winnow is an abstract \texttt{CalibrationFeatures} class, which users can subclass to implement custom feature computations.
The system includes a set of default features that rely on metadata such as predicted peptide sequences, mass spectra outputs, retention times, precursor masses and beam search results.
Custom features can access these same inputs, but are not limited to them; users can introduce new data sources or experimental metadata as needed.
This flexibility allows researchers to incorporate domain-specific information that may improve the calibration model's performance for their particular experimental conditions or sample characteristics.

The modular design of our calibration system allows features to be easily added or removed from the pipeline.
For example, the Prosit-based features (\texttt{PrositFeatures} and \texttt{RetentionTimeFeature}) rely on external models with restrictions on supported PTMs, precursor charges and sequence lengths.
When these constraints arise, these features can simply be omitted without disrupting the calibration pipeline, allowing users to adapt the feature set to their data rather than conforming their data to the model.
Additionally, Winnow outputs all computed features as part of its dataset metadata, supporting transparency and enabling downstream analysis to identify which features are most informative for a given dataset.
These design choices provide researchers with openness, adaptability and long-term utility, ensuring that Winnow is useful across a wide range of proteomics applications and evolving experimental protocols.

\section{Discussion}

In this study, we present Winnow, a model-agnostic calibration framework for estimating false discovery rates in DNS. 
Utilising features derived from both model outputs and experimental spectra, Winnow provides calibrated confidence scores alongside robust statistical metrics such as FDR, q-values and posterior error probabilities. 
Our results demonstrate that Winnow accurately and consistently estimates these metrics across diverse datasets.

This framework immediately enhances the usability and reliability of DNS results. 
By enabling probabilistic scoring and rigorous error estimation, Winnow aligns DNS workflows with established database search standards. 
It offers practical tools for calibration and filtering, including default thresholds and interpretable metrics that streamline downstream analysis and facilitate model evaluation in discovery settings.

Although database-grounded extrapolation may perform well on certain datasets, its reliability hinges on the assumption that database-labelled spectra and spectra that failed to obtain a database label share similar characteristics and thus similar error profiles and confidence score distributions.
This assumption can hold in well-characterised datasets, where the proteome is highly complete and many real peptides receive labels.
However, it breaks down in more typical DNS contexts involving under-characterised organisms, sequence variants or mixed-species samples.
In these cases, the labelled subset tends to reflect only a narrow and closest-to-ideal slice of the spectrum landscape (e.g., primarily well-known and easily identifiable peptides) while the unlabelled set includes a broader range of more challenging spectra.
As a result, extrapolating thresholds from the labelled data can lead to underestimated FDR and unreliable identifications.
In contrast, Winnow does not rely on the presence or representativeness of labelled data, enabling more robust FDR control.

Nonetheless, several limitations remain.
Although designed to be data-agnostic, our current implementation is trained using outputs from InstaNovo and limited to Orbitrap-derived spectra. 
This may introduce biases and limit generalisability to other DNS models, instrument types or acquisition methods.
Future iterations will address this by incorporating more diverse training data. 
Although still improved relative to raw DNS model confidence, the held-out Immunopeptidomics-2 dataset exhibited poorer calibration and, consequently, less accurate non-parametric FDR estimation.
This likely reflects the scarcity of non-tryptic spectra--prevalent in this dataset--which constituted less than 1\% of our training data.
Addressing this imbalance will be important for future model releases.
Additionally, our reliance on Prosit-derived features (e.g., iRT and fragment intensity) restricts the framework to peptides and modifications supported by Prosit. 
While these features can be omitted, enabling broader applicability, extending predictive models to cover more modifications would further strengthen Winnow’s utility.

Beam search features, particularly margin between best and second best sequence prediction, emerged as strong predictors of correctness. 
This was expected, as a wide margin typically signals high model confidence. 
What is notable, however, is the dominance of this feature over experimental ones. 
This highlights the strength of sequence-intrinsic signals, but also points to a caveat: beam scores are sometimes uniformly low, and when fewer than three predictions are returned, zero-padding may artificially inflate variance-based features. 
Addressing these cases is a priority for future development.

We also expected retention time to play a more prominent role in detecting false positives. 
Its limited contribution may reflect suboptimal mapping between observed RT and predicted iRT, or it may suggest that false positives resemble true sequences closely enough to retain plausible chromatographic behaviour. 
It’s also possible that DNS models implicitly learn to favour peptides with realistic physicochemical properties, even when predictions are incorrect.

Chimeric spectra remain a significant challenge in DNS. 
To partially address this, we incorporated features from second-best predictions, reasoning that elevated confidence and ion match rates in both top and runner-up sequences might indicate spectral ambiguity. 
Surprisingly, we found that correct PSMs often possessed moderate levels of chimeric ion match intensity, whereas incorrect PSMs were more frequently associated with low chimeric ion intensity values.
This suggests that partial secondary matches can coexist with confident identifications, reflecting genuine fragment sharing.
Such cases may arise when spectra possess high fragment ion coverage and resemble examples commonly seen during training.
By contrast, an absence of secondary matches may instead signal noise or novel peptides, where the DNS model lacks credible runner-up candidates and correspondingly assigns low confidence.
Although Winnow does not capture chimericity in the manner we initially expected, we found these features to provide context-dependent signal that improves calibration.

Our ground-truth assignments rely on database search matches, which are widely considered to be accurate but still imperfect. 
We also use direct proteome mapping as an alternative or complementary method of assess PSM correctness.
Even when a peptide maps to the reference proteome, this does not guarantee it is the correct match for the observed spectrum.
Conversely, truly correct sequences may fail to map to the reference at all, particularly in samples with high proteome novelty.
Thus, database search-based labelling, and especially proteome mapping, should be viewed only as a proxy for correctness rather than a definitive truth.

In practice, the probability of a correct peptide identification may vary between datasets--a phenomenon known as label shift.
This violates the key assumption in classification models that the class priors (i.e., the proportions of correct and incorrect PSMs) are stable between training and test data.
A model may systematically over- or under-estimate the probability that a given PSM is correct, reducing calibrator performance and potentially degrading downstream FDR estimation.
In future work, it will be important to investigate approaches that explicitly account for changing class priors, for example by incorporating mechanisms to recalibrate model outputs when applied to new datasets.
Such strategies would help ensure that probability estimates remain reliable even under label shift, thereby strengthening the robustness of FDR estimation across diverse experimental conditions.

Beyond label shift, feature shift also limits generalisation.
While Winnow performs well on data from our own lab during our hold-one-out analysis, its performance decreases on external datasets, likely because shared instrumentation among our validation sets masks distributional differences.
Although zero-shot application is possible, dataset-specific calibration or retraining is advised for optimal results. 
We are actively expanding our training set across instruments and labs to address this.

Recent work by Sanders et al. proposed a complementary procedure for FDR control in DNS that uses database search results to model correct score distributions \cite{sanders_procedure_2025}. 
Winnow is designed to operate in a fully \textit{de novo} context, without requiring such external anchors.
Instead of making assumptions about class-conditional distributions, Winnow learns to directly model the conditional probability of correctness given DNS model outputs and supplementary calibration features.

Looking ahead, Winnow’s capabilities could be improved by incorporating additional calibration features and allowing greater flexibility in calibration model hyperparameters.
Beyond this, exploring alternative architectures such as linear discriminant analysis separators or transformers may further improve performance. 
Equally important is the use of more diverse datasets spanning instruments, species and peptide chemistries, which will be key to achieving broader generalisation.

A promising application of Winnow lies in hybrid workflows that combine DNS with multiple approaches or database search. 
Calibrated scores from Winnow could be used for integration, filtering and ensembling across models and engines. 
This could enable joint analyses, improving both peptide recovery and confidence. 
Whether a unified calibration model can accommodate predictions from diverse tools, or whether tool-specific calibrations are needed, remains an open question.

Ultimately, Winnow enhances confidence and interpretability in DNS, especially as more diverse training data are utilised. 
Future versions can broaden support for instruments and models, and users working beyond the Prosit-compatible space can retrain Winnow using available features.
By providing calibrated scores and principled FDR control, Winnow bridges a longstanding gap in DNS pipelines. 
It enables statistical evaluation of DNS results and integration with database searches, unlocking more accurate, scalable and interpretable workflows in proteomic discovery.

\section{Methods}

\subsection{Datasets}

We used ten publicly available datasets in this study, comprising a total of 3,691,384 spectra (Supplementary Fig. \ref{fig:dataset_characteristics}A).
These datasets span a diverse range of organisms and experimental contexts, supporting evaluation across both standard and challenging peptide sequencing scenarios.

Approximately half of our data is derived from human samples including HeLa Single Shot (17,683 PSMs from 46,409 spectra); HepG2 (292,736 PSMs from 1,259,452 spectra); Wound Exudates (3,727 PSMs from 105,591 spectra); Herceptin, a monoclonal antibody dataset (804 PSMs from 58,609 spectra); and two immunopeptidomics datasets that reflect HLA-presented peptides: one with 652 PSMs from 404,062 spectra (referred to as Immunopeptidomics-1) and another with 7,594 PSMs from 100,853 spectra (Immunopeptidomics-2).

To support model generalisation beyond human samples, we included datasets from \textit{C. elegans} (232,156 PSMs from 883,187 spectra), a well-studied multicellular organism; \textit{Candidatus} Scalindua Brodae (9,053 PSMs from 26,103 spectra), a marine bacterial species; and Snake Venomics (3,727 PSMs from 600,955 spectra) comprising spectra from multiple venomous species with highly diverse and often under-annotated proteomes. 
The HeLa Degradome dataset (113,373 PSMs from 206,163 spectra), although also human-derived, offers further variation by providing peptides created using a short incubation with the gluC protease to generate neo-N-termini before complete tryptic digestion, resulting in semi-tryptic peptides and a degradomic profile.

The datasets HeLa Single Shot, Wound Exudates, Herceptin, Immunopeptidomics-1, Scalindua Brodae, Snake Venomics, HeLa Degradome and HepG2 were used to train our general calibrator model, with the combined data split 90/10/10 into training, validation, and an in-distribution test set. 
To further assess model generalisation, the datasets \textit{C. elegans} and Immunopeptidomics-2 were held out entirely for evaluation. 
All datasets were obtained from publicly available repositories, with accession numbers and references provided in Section \ref{sec:data_availability}.

\subsubsection{Data pre-processing}

All spectra were preprocessed to retain only those with precursor charges below +6. 
We apply additional filtering for spectra associated with PSMs: removing sequences containing the amino acids selenocysteine, pyrrolysine or an unknown amino acid (denoted by `X') and any sequences with modifications other than methionine oxidation. 
Additionally, all cysteines were treated as carbamidomethylated.
Each spectrum was annotated with predicted peptide sequences using InstaNovo\cite{eloff_instanovo_2025} (v1.1.1), employing knapsack beam search with a maximum beam length of 50.
Spectra were discarded if InstaNovo failed to produce any candidate sequences or if the first- or second-ranked predictions exceed 30 amino acids.

These thresholds were selected to align with the assumptions of our calibrator model, particularly those originating from the use of Prosit features (iRT and fragment intensity predictions), which impose length and sequence constraints.

\subsection{Confidence score calibration}

In Winnow, calibration serves multiple purposes. 
It enables the use of arbitrary real-valued confidence scores, and provides a unified framework that includes common database search post-processing steps such as incorporation of additional features and PSM rescoring. 
Most importantly, it ensures that the resulting probabilities satisfy the calibration assumption of our novel FDR estimator.
At the same time, calibration can improve separability between correct and incorrect PSMs, ideally yielding confidences close to 1 for true matches and 0 for false ones, which would further strengthen downstream error control.

\subsubsection{Motivation and formulation}

Not all peptide sequencing methods return confidence scores interpretable as probabilities. 
We address this by fitting a calibration model that outputs the estimated probability that a PSM identification is correct.

Let $\mathcal{D}_C = \{(\mathbf{x}_i, y_i)\}_{1\leq i \leq n}$ denote a dataset of gold-standard PSMs, where $\mathbf{x}_i \in \mathcal{X}_S$ is the $i$-th spectrum and $y_i \in \mathcal{Y}_P$ the $i$-th gold-standard peptide, and $\mathcal{X}_S$ and $\mathcal{Y}_P$ are the sets of possible spectra and peptides, respectively. 
A peptide sequencing method is defined as a function $f: \mathcal{X}_S \rightarrow \mathcal{Y}_P\times \mathbb{R}$ that maps a spectrum $\mathbf{x}$ to a predicted peptide and confidence score: $f(\mathbf{x}) = (\hat{y}, S)$.

From peptide sequencing output, we derive a new dataset of spectra, confidence scores and correctness labels $(\mathbf{x}_i, S_i, I(y_i = \hat{y}_i))$ and train a supervised classifier to estimate $\hat S_i = P(y_i = \hat{y}_i \mid S_i, \phi(\mathbf{x}_i))$, where $\phi(\mathbf{x}_i)$ denotes optional features such as iRT error, mass shift or precursor charge. 
Flexibility in the choice of calibrator means the calibration step subsumes traditional rescoring and model combination as special cases.

As part of score calibration, raw model probabilities are transformed to reflect true probabilities.
This probability calibration is crucial for the reliable estimation of downstream statistical measures such as FDR and PEP, and it expands Winnow's compatibility to a diverse range of scoring strategies.

\subsubsection{Calibration features}

To improve the reliability of peptide identification confidence scores, we developed a set of calibration features that supplement raw model confidence scores.
These features were selected, based on literature and our assessment, to provide insights on the quality of the match, using differences in theoretical and experimental spectra and the confidence of the model in its predictions.

\textit{\textbf{Mass accuracy and retention time features}}
We calculated the mass error as the difference between the observed precursor mass and the theoretical mass of the identified peptide, accounting for the mass of water and a proton.
This feature helps identify potential misidentifications by detecting mass deviations, which can arise from incorrect sequence assignments or modifications.
For retention time information, we used Prosit's \cite{gessulat_prosit_2019} iRT predictions to create training labels, predicting iRT \cite{escher_using_2012} values for the top ten percent of our high-confidence sequences.
We then trained a multi-layer perceptron (MLP) to map observed retention times to predicted iRT values. 
Finally, we computed the absolute error between predicted and actual (that is, Prosit-predicted) iRT values.
This approach accounted for variations in retention time behaviour between different runs while maintaining the relative ordering of peptides, without spike-in peptides.
The iRT error provides an additional dimension of validation, because peptides that elute significantly earlier or later than predicted may indicate incorrect identifications or unexpected modifications.

\textit{\textbf{Spectral matching features}}
We computed the fraction of theoretical fragment ions that match experimental peaks within a specific mass tolerance, along with their corresponding intensities.
These features are derived by comparing theoretical fragment ion m/z values predicted by Prosit against the experimental spectrum.
The ion match rate, which represents the proportion of theoretical ions that find a match in the experimental spectrum, provides a measure of spectral similarity.
The match intensity, which captures the relative abundance of matched peaks, offers complementary information by distinguishing between strong and weak spectral matches.
For example, a high match rate with low intensities might indicate a correct but weak identification, while a high match rate with high intensities suggests a strong, confident identification.
To assess potential chimeric spectra--spectra containing fragment ions from multiple peptides--we computed similar matching metrics for the second-best peptide sequence from the beam search.
High chimeric match rates or intensities may indicate the presence of multiple peptides in the spectrum, which could affect the reliability of the identification.

\textit{\textbf{Sequence prediction features}}
We quantified the confidence of the top-ranked sequence through several complementary metrics that capture different aspects of prediction uncertainty (Fig. \ref{fig:winnow_overview}D).
The margin measures the probability gap between the top-ranked and second-ranked sequence (often referred to as the nextscore in database search engines), while the median margin represents the difference between the top sequence and the median probability of all runner-up sequences. 
These metrics help identify cases where the model is uncertain between multiple plausible sequences. 
To further quantify uncertainty over the sequence label distribution for a given spectrum, we computed the Shannon entropy of the normalised probabilities of runner-up sequences from the beam search.
Finally, we calculated a z-score that measures how many standard deviations the top beam score lies from the mean of all beam search results for a given spectrum, helping to identify unusually strong or weak predictions relative to the candidate pool.
Together, these features provided further, orthogonal, validation to our mass accuracy, retention time and spectral matching features.

\subsubsection{Calibration method}

Winnow calibrates raw model confidence scores using a neural network.
Specifically, Winnow's calibrator constituted an MLP binary classifier with two hidden layers of 50 units each, trained via cross-entropy loss with L2 regularisation (regularisation coefficient $\alpha=0.0001$) and a learning rate of 0.0001.
Input features were first standardised to zero mean and unit variance before being passed into the network.
The MLP learned a mapping from raw confidence score and optional contextual features $\phi(\mathbf{x})$ to a calibrated probability estimate $\hat S_i = P(C = 1 \mid S_i, \phi(\mathbf{x}_i)$, where $C \in \{0, 1\}$ indicates whether a PSM is correct.
We used early stopping, holding out ten percent of the training data, to combat overfitting.

\subsection{FDR estimation}

FDR estimation is essential in MS-based peptide identification to quantify the expected proportion of misidentifications among accepted PSMs; it is the expected proportion of false positives among all accepted predictions.
While target-decoy strategies are commonly used in database searches, alternative approaches are needed for DNS methods.
Formally, for a confidence threshold $\tau \in [0, 1]$, the FDR can be defined as
\begin{equation}
    \text{FDR}(\tau) = P(C=0 \mid S \ge \tau), \label{eqn:fdr_definition}
\end{equation}
where $C$ is a binary indicator representing the correctness of the identification and $S$ is the model confidence score.
Therefore, FDR is the probability that a prediction is incorrect given that its probability score exceeds $\tau$.

Winnow introduces a novel approach to FDR control in PSM analysis, while maintaining compatibility with established methods for FDR control.
Our pipeline's primary innovation lies in its non-parametric FDR control method, which solves the challenge of FDR estimation in DNS.

Winnow's framework provides comprehensive PSM quality assessment through three complementary metrics: 
\begin{enumerate}
    \item \textbf{Experiment-wide FDR control}
    \begin{description}
        \item[] This maintains a target error rate across the entire dataset. 
    \end{description}
    \item \textbf{Q-values}
    \begin{description}
        \item[] These enable granular evaluation of individual PSMs, indicating the minimum FDR threshold at which a given PSM would be considered significant.
    \end{description}
    \item \textbf{Posterior error probabilities (PEP)}.
    \begin{description}
        \item[] These provide direct estimates of the probability that each individual PSM is correct.
    \end{description}
\end{enumerate}
Together, these metrics offer a flexible validation strategy that combines experiment-wide error control with fine-grained PSM-specific interpretability.

\subsubsection{Database-grounded FDR estimation}

Estimating FDR in DNS is challenging due to the absence of decoy sequences and target databases.
We address this by grounding our FDR estimates in results from a conventional target-decoy database search conducted on the same spectra.
The procedure is as follows:
\begin{enumerate}
    \item \textbf{Reference PSMs from database search}
    \begin{description}
        \item[] A standard database search is first applied to the spectra, producing PSMs. These are treated as reference labels for FDR estimation.
    \end{description}
    \item \textbf{Scoring DNS predictions}
    \begin{description}
        \item[] The DNS model is then applied to the same spectra, outputting a log-probability for each predicted peptide sequence. The log-probabilities are exponentiated to yield confidence scores in the range of $[0, 1]$. These model probabilities may be optionally calibrated using Winnow.
    \end{description}
    \item \textbf{Alignment and labelling}
    \begin{description}
        \item[] Scan indices are aligned between the database results and DNS predictions. Each prediction is labelled as a true positive (TP) if the predicted peptide matches the reference peptide and a false positive (FP) otherwise. Matching accounts for minor mass shifts and post-translational modifications.
    \end{description}
    \item \textbf{Thresholding by confidence score}
    \begin{description}
        \item[] The predictions are sorted in descending order of confidence. For any confidence $s \in [0,1]$, we compute the precision threshold of $s$ as,
        \begin{equation}
            \text{Precision}(s) = \frac{\sum_{i=1}^{n} C_i \cdot I(S_i \geq s)}{\sum_{i=1}^{n} I(S_i \geq s)},
        \end{equation}
        where $C_i$ is a binary indicator representing the correctness of the predicted sequence.
        We then estimate FDR as
        \begin{equation}
            \begin{split}
                \text{FDR}(\tau) &= 1 - \text{Precision}(\tau)\\
                        &= 1 - \frac{\sum_{i=1}^{n} C_i \cdot I(S_i \geq \tau)}{\sum_{i=1}^{n} I(S_i \geq \tau)}\\
                        &= \frac{\sum_{i=1}^{n} I(S_i \geq \tau) - \sum_{i=1}^{n} C_i \cdot I(S_i \geq \tau)}{\sum_{i=1}^{n} I(S_i \geq \tau)}\\
                        &= \frac{\sum_{i=1}^{n}(1 - C_i) \cdot I(S_i \geq \tau)}{\sum_{i=1}^{n} I(S_i \geq \tau)}
            \end{split}
        \end{equation}
        A confidence cutoff $\tau$ is then found such that $\text{FDR}(\tau) \leq \alpha$, for a chosen FDR threshold $\alpha$ (typically 5\%).
    \end{description}
    \item \textbf{Filtering and downstream use}
    \begin{description}
        \item[] The threshold $\tau$ is applied to unlabelled predictions. Those with scores above $\tau$ are retained for further use; those below are discarded.
    \end{description}
    \item \textbf{Reporting novel sequences}
    \begin{description}
        \item[] Among the high-confidence retained predictions, we report the proportion that match database identifications and the proportion that may represent novel peptide sequences.
    \end{description}
\end{enumerate}

This method provides an empirical way to control FDR in DNS, leveraging the high-confidence subset of database search results as a proxy ground truth.
However, it is reliant on the availability of reference databases and alignment between search and model outputs, and the threshold it yields may not generalise well to DNS predictions outside the database space.

\subsubsection{Non-parametric FDR estimation}

In contrast, Winnow's novel non-parametric FDR estimator operates directly on confidence scores without necessitating gold-standard database matches. 

The formulation of FDR in \ref{eqn:fdr_definition} can be broken down using the definition of conditional probability into
\begin{equation}
    \text{FDR}(\tau) = \frac{P(C = 0 , S \ge \tau)}{P(S \ge \tau)}. \label{eqn:fdr_bayes_decomposition}
\end{equation}
Prior work \cite{keller_empirical_2002, gonnelli_decoy-free_2015, peng_algorithm_2024, peng_new_2020, madej_modeling_2023} has further decomposed the numerator in \ref{eqn:fdr_bayes_decomposition} in a generative fashion as
\begin{equation}
    P(C = 0 , S \ge \tau) = \int_{\tau}^{1} P(S = s \mid C = 0) \cdot P(C = 0) ds,  \label{eqn:fdr_numerator}
\end{equation}
where $P(S|C=0)$ is a learned negative or `decoy' distribution.

This method of FDR estimation requires correctly specifying the class-conditional distributions ($P(S|C=0)$ and $P(S|C=1)$) as well as learning their parameters and the mixture weight $P(C=1)$ from unlabelled scores. 
The correct model specification will vary from one sequencing method and dataset to the next and is difficult to verify, and misspecification will lead to inaccurate FDR estimates. 
Further, even when the model is correctly specified, the parameters may be poorly identified leading to unstable and inaccurate FDR estimation. 

However, it is also possible to use a discriminative decomposition of the numerator in \ref{eqn:fdr_bayes_decomposition}:
\begin{equation}
    P(C = 0 , S \ge \tau) = \int_{\tau}^{1} P(C = 0 \mid S = s) \cdot P(S = s) ds, \label{eqn:discriminative_fdr}
\end{equation}
which, as we show below, allows us to estimate FDR without requiring a parametric distribution over scores.

We can write $P(S \ge \tau)$ from \ref{eqn:fdr_bayes_decomposition} as
\begin{equation}
    P(S \ge \tau) = \int_{\tau}^{1} P(S = s)ds. \label{eqn:fdr_denominator}
\end{equation}
Substituting \ref{eqn:fdr_numerator} and \ref{eqn:fdr_denominator} into \ref{eqn:fdr_bayes_decomposition}, we obtain
\begin{equation}
    \text{FDR}(\tau) = \frac{\int_{\tau}^{1}P(C = 0 \mid S = s) \cdot P(S = s) ds}{\int_{\tau}^{1} P(S = s) ds}. \label{eqn:fdr_expanded_no_calibration}
\end{equation}

To proceed, we will assume that our confidence scores are probabilities and that they are \emph{calibrated} (i.e., the probability of a PSM being correct is equal to its confidence score).

Formally, a probability estimator is calibrated when the following holds:
\begin{equation}
    P(C = 1 | S = s) = s, \label{eqn: calibration}
\end{equation}
where the PSM confidence score $S$ is a probability and $C$ is an indicator for whether the identification is correct.
Equivalently, we can write
\begin{equation}
    P(C = 0 | S = s) = 1 - s. \label{eqn: calibration_negation}
\end{equation}

Assuming calibration, we can substitute \ref{eqn: calibration_negation} into \ref{eqn:fdr_expanded_no_calibration} to obtain
\begin{equation}
    \text{FDR}(\tau) = \frac{\int_{\tau}^{1} (1 - s) \cdot P(S = s) ds}{\int_{\tau}^{1} P(S = s) ds}.
\end{equation}

For a given confidence threshold $\tau$, the FDR may then be estimated non-parametrically as
\begin{equation}
    \widehat{\text{FDR}}(\tau) = \frac{\sum_{i=1}^{n}(1 - s_i) \cdot I(s_i \ge \tau)}{\sum_{i=1}^{n}I(s_i \ge \tau)},  \label{eqn:non_parametric_estimator}
\end{equation}
where $s_i$ represents the calibrated confidence score for the $i$-th PSM for $i \in \{1, \ldots, N\}$.
This approach offers computational efficiency and robustness across various experimental conditions, making it particularly valuable when reference databases are incomplete or when rapid analysis is needed, requiring only well-calibrated PSM confidence scores.

Winnow's empirical FDR estimation procedure is as follows:
\begin{enumerate}
    \item \textbf{Scoring DNS predictions}
    \begin{description}
        \item[] We proceed similarly to the database-grounded approach, obtaining model confidence scores $S_i \in [0,1]$ for each predicted peptide sequence. These scores are assumed to be calibrated, meaning they directly represent the probability of a correct prediction.
    \end{description}
    \item \textbf{Error probability computation}
    \begin{description}
        \item[] For each confidence score $S_i$, we compute the error probability as $E_i = 1 - S_i$. This represents the probability of an incorrect prediction at each confidence level.
    \end{description}
    \item \textbf{Cumulative error estimation}
    \begin{description}
        \item[] The predictions are sorted in descending order of confidence. For any confidence threshold $s$, we compute the cumulative error probability as
    \begin{equation}
    \text{CumError}(s) = \sum_{i=1}^{n} E_i \cdot I(S_i \geq s),
    \end{equation}
    where $I(S_i \geq s)$ is an indicator function for scores above the threshold. The number of predictions above the threshold is
    \begin{equation}
    N(s) = \sum_{i=1}^{n} I(S_i \geq s).
    \end{equation}
    \end{description}
    \item \textbf{FDR estimation}
    \begin{description}
        \item[] The FDR at confidence threshold $s$ is estimated as the ratio of cumulative errors to the number of predictions:
    \begin{equation}
    \text{FDR}(s) = \frac{\text{CumError}(s)}{N(s)}.
    \end{equation}
    This provides a non-parametric estimate of the false discovery rate that makes no assumptions about the underlying distribution of scores.
    \end{description}
    \item \textbf{Thresholding by confidence score}
    \begin{description}
        \item[] For a chosen FDR threshold $\alpha$ (typically 5\%), we find the smallest confidence score $\tau$ such that $\text{FDR}(\tau) \leq \alpha$.
    \end{description}
    \item \textbf{Filtering and downstream use}
    \begin{description}
        \item[] The threshold $\tau$ is applied to filter predictions. Those with scores above $\tau$ are retained for further analysis, while those below are discarded.
    \end{description}
\end{enumerate}

\subsection{PSM-specific FDR metrics}

In addition to experiment-wide FDR control, Winnow provides fine-grained quality assessment through the reporting of PSM-specific FDR metrics that can be used for further filtering (Fig. \ref{fig:winnow_overview}C).

\subsubsection{Q-values}

A q-value provides individual FDR estimates for each PSM based on its confidence score.
This is achieved by computing the least conservative FDR that would be obtained if using that PSM's confidence score as a threshold.
This local approach allows the user to analyse and make decisions about each PSM while maintaining awareness of their contribution to the overall FDR.

\subsubsection{Posterior error probabilities (PEP)}

Posterior error probabilities provide an additional perspective to FDR estimates by offering direct, PSM-specific error estimates.
For a given PSM confidence score $s$, PEP is defined as
\begin{equation}
    \text{PEP}(s) = P(C = 0 | S = s).
\end{equation}
Given our definition of calibration in \ref{eqn: calibration_negation}, PEP simply becomes
\begin{equation}
    \text{PEP}(s) = P(C = 0 | S = s) = 1 - s.
\end{equation}
PEP provides an immediate assessment of individual PSM reliability, enabling users to make decisions about individual identifications without waiting for sufficient data for empirical FDR estimation.

\section{Data availability} \label{sec:data_availability}

We utilised ten different datasets in this study. 
The Single Shot HeLa proteome, HeLa Degradome and \textit{Candidatus} Scalindua Brodae raw data and search results were obtained from the InstaNovo study and are deposited in the PRIDE \cite{perezriverol_pride_2021} repository with dataset identifier PXD044934.
The Herceptin dataset is available on figshare at \url{https://doi.org/10.6084/m9.figshare.21394143} \cite{beslic_comprehensive_2023}.
The Snake Venomics dataset and search results can be found in the PRIDE repository with identifier PXD036161 \cite{nguyen_highthroughput_2022}.
The Wound Exudates dataset is available through PanoramaWeb with dataset identifier PXD025748 \cite{mikosinski_longitudinal_2022}. 
The HepG2 and \textit{C. elegans} datasets were retrieved from a study on the proteome of different kingdoms of life\cite{muller_proteome_2020} and are available from the PRIDE repository with identifier PXD019483 and PXD014877.
The Immunopeptidomics-1 dataset can be found in the PRIDE repository with identifier PXD006939 \cite{chong_highthroughput_2018}.
The Immunopeptidomics-2 dataset was retrieved from the PRIDE repository with dataset identifier PXD023064.
All datasets are additionally available on Hugging Face at \url{https://huggingface.co/datasets/InstaDeepAI/winnow-ms-datasets}.
The Winnow outputs have been deposited to Figshare and are can be accessed with the link \url{10.6084/m9.figshare.30147601}.

Model checkpoints are available on Hugging Face at \url{https://huggingface.co/InstaDeepAI/winnow-general-model} and \url{https://huggingface.co/InstaDeepAI/winnow-helaqc-model}.
Models resulting from hold-out-out analysis are accessible at \url{10.6084/m9.figshare.30147364} (Fig. \ref{fig:general_calibrator_perf_and_features}A).

\section{Code availability}

Winnow is available at \url{https://github.com/instadeepai/winnow}.
Our code includes usage documentation and a user-friendly command line interface.
We provide a Google Colab notebook that introduces Winnow and makes the model easily accessible to users.
Jupyter notebooks to reproduce our figures can be found on Figshare at \url{10.6084/m9.figshare.30147472}, and extra scripts for feature importance and hold-one-out analysis are at \url{10.6084/m9.figshare.30147463}.

\bibliographystyle{unsrt}
\bibliography{references}

\begin{thebibliography}{10}

\bibitem{aebersold_mass-spectrometric_2016}
Ruedi Aebersold and Matthias Mann.
\newblock Mass-spectrometric exploration of proteome structure and function.
\newblock {\em Nature}, 537(7620):347--355, September 2016.

\bibitem{sadygov_large-scale_2004}
Rovshan~G. Sadygov, Daniel Cociorva, and John R.~3rd Yates.
\newblock Large-scale database searching using tandem mass spectra: looking up the answer in the back of the book.
\newblock {\em Nature methods}, 1(3):195--202, December 2004.
\newblock Place: United States.

\bibitem{elias_target-decoy_2010}
Joshua~E. Elias and Steven~P. Gygi.
\newblock Target-decoy search strategy for mass spectrometry-based proteomics.
\newblock {\em Methods in molecular biology (Clifton, N.J.)}, 604:55--71, 2010.
\newblock Place: United States.

\bibitem{chick_mass-tolerant_2015}
Joel~M. Chick, Deepak Kolippakkam, David~P. Nusinow, Bo~Zhai, Ramin Rad, Edward~L. Huttlin, and Steven~P. Gygi.
\newblock A mass-tolerant database search identifies a large proportion of unassigned spectra in shotgun proteomics as modified peptides.
\newblock {\em Nature biotechnology}, 33(7):743--749, July 2015.
\newblock Place: United States.

\bibitem{savitski_scalable_2015}
Mikhail~M. Savitski, Mathias Wilhelm, Hannes Hahne, Bernhard Kuster, and Marcus Bantscheff.
\newblock A {Scalable} {Approach} for {Protein} {False} {Discovery} {Rate} {Estimation} in {Large} {Proteomic} {Data} {Sets}[{S}].
\newblock {\em Molecular \& Cellular Proteomics}, 14(9):2394--2404, 2015.

\bibitem{freestone_re-investigating_2023}
Jack Freestone, William~Stafford Noble, and Uri Keich.
\newblock Re-investigating the correctness of decoy-based false discovery rate control in proteomics tandem mass spectrometry.
\newblock {\em bioRxiv}, 2023.
\newblock Publisher: Cold Spring Harbor Laboratory \_eprint: https://www.biorxiv.org/content/early/2023/06/24/2023.06.21.546013.full.pdf.

\bibitem{wen_assessment_2025}
Bo~Wen, Jack Freestone, Michael Riffle, Michael~J. MacCoss, William~S. Noble, and Uri Keich.
\newblock Assessment of false discovery rate control in tandem mass spectrometry analysis using entrapment.
\newblock {\em Nature Methods}, June 2025.

\bibitem{keller_empirical_2002}
Andrew Keller, Alexey~I. Nesvizhskii, Eugene Kolker, and Ruedi Aebersold.
\newblock Empirical {Statistical} {Model} {To} {Estimate} the {Accuracy} of {Peptide} {Identifications} {Made} by {MS}/{MS} and {Database} {Search}.
\newblock {\em Analytical Chemistry}, 74(20):5383--5392, October 2002.
\newblock Publisher: American Chemical Society.

\bibitem{gonnelli_decoy-free_2015}
Giulia Gonnelli, Michiel Stock, Jan Verwaeren, Davy Maddelein, Bernard De~Baets, Lennart Martens, and Sven Degroeve.
\newblock A {Decoy}-{Free} {Approach} to the {Identification} of {Peptides}.
\newblock {\em Journal of Proteome Research}, 14(4):1792--1798, April 2015.
\newblock Publisher: American Chemical Society.

\bibitem{peng_algorithm_2024}
Yisu Peng, Shantanu Jain, and Predrag Radivojac.
\newblock An algorithm for decoy-free false discovery rate estimation in {XL}-{MS}/{MS} proteomics.
\newblock {\em Bioinformatics}, 40(Supplement\_1):i428--i436, July 2024.

\bibitem{peng_new_2020}
Yisu Peng, Shantanu Jain, Yong~Fuga Li, Michal Greguš, Alexander~R. Ivanov, Olga Vitek, and Predrag Radivojac.
\newblock New mixture models for decoy-free false discovery rate estimation in mass spectrometry proteomics.
\newblock {\em Bioinformatics (Oxford, England)}, 36(Suppl\_2):i745--i753, December 2020.
\newblock Place: England.

\bibitem{madej_modeling_2023}
Dominik Madej and Henry Lam.
\newblock Modeling {Lower}-{Order} {Statistics} to {Enable} {Decoy}-{Free} {FDR} {Estimation} in {Proteomics}.
\newblock {\em Journal of proteome research}, 22(4):1159--1171, April 2023.
\newblock Place: United States.

\bibitem{huang_development_2020}
Jiangming Huang, Biyun Jiang, Huanhuan Zhao, Mengxi Wu, Siyuan Kong, Mingqi Liu, Pengyuan Yang, and Weiqian Cao.
\newblock Development of a {Computational} {Tool} for {Automated} {Interpretation} of {Intact} {O}-{Glycopeptide} {Tandem} {Mass} {Spectra} from {Single} {Proteins}.
\newblock {\em Analytical Chemistry}, 92(9):6777--6784, May 2020.
\newblock Publisher: American Chemical Society.

\bibitem{ma_peaks_2003}
Bin Ma, Kaizhong Zhang, Christopher Hendrie, Chengzhi Liang, Ming Li, Amanda Doherty-Kirby, and Gilles Lajoie.
\newblock {PEAKS}: powerful software for peptide de novo sequencing by tandem mass spectrometry.
\newblock {\em Rapid communications in mass spectrometry : RCM}, 17(20):2337--2342, 2003.
\newblock Place: England.

\bibitem{frank_pepnovo_2005}
Ari Frank and Pavel Pevzner.
\newblock {PepNovo}: {De} {Novo} {Peptide} {Sequencing} via {Probabilistic} {Network} {Modeling}.
\newblock {\em Analytical Chemistry}, 77(4):964--973, February 2005.

\bibitem{muth_evaluating_2018}
Thilo Muth and Bernhard~Y Renard.
\newblock Evaluating de novo sequencing in proteomics: already an accurate alternative to database-driven peptide identification?
\newblock {\em Briefings in Bioinformatics}, 19(5):954--970, September 2018.

\bibitem{muth_potential_2018}
Thilo Muth, Felix Hartkopf, Marc Vaudel, and Bernhard~Y. Renard.
\newblock A {Potential} {Golden} {Age} to {Come}-{Current} {Tools}, {Recent} {Use} {Cases}, and {Future} {Avenues} for {De} {Novo} {Sequencing} in {Proteomics}.
\newblock {\em Proteomics}, 18(18):e1700150, September 2018.
\newblock Place: Germany.

\bibitem{kall_posterior_2008}
Lukas Käll, John~D. Storey, Michael~J. MacCoss, and William~Stafford Noble.
\newblock Posterior {Error} {Probabilities} and {False} {Discovery} {Rates}: {Two} {Sides} of the {Same} {Coin}.
\newblock {\em Journal of Proteome Research}, 7(1):40--44, January 2008.
\newblock Publisher: American Chemical Society.

\bibitem{qiao_computationally_2021}
Rui Qiao, Ngoc~Hieu Tran, Lei Xin, Xin Chen, Ming Li, Baozhen Shan, and Ali Ghodsi.
\newblock Computationally instrument-resolution-independent de novo peptide sequencing for high-resolution devices.
\newblock {\em Nature Machine Intelligence}, 3(5):420--425, May 2021.

\bibitem{mao_mitigating_2023}
Zeping Mao, Ruixue Zhang, Lei Xin, and Ming Li.
\newblock Mitigating the missing-fragmentation problem in de novo peptide sequencing with a two-stage graph-based deep learning model.
\newblock {\em Nature Machine Intelligence}, 5(11):1250--1260, November 2023.

\bibitem{yilmaz_sequence--sequence_2024}
Melih Yilmaz, William~E. Fondrie, Wout Bittremieux, Carlo~F. Melendez, Rowan Nelson, Varun Ananth, Sewoong Oh, and William~Stafford Noble.
\newblock Sequence-to-sequence translation from mass spectra to peptides with a transformer model.
\newblock {\em Nature Communications}, 15(1):6427, July 2024.

\bibitem{zhao_transformer-based_2025}
Yang Zhao, Shuo Wang, Jinze Huang, Bo~Meng, Dong An, Xiang Fang, Yaoguang wei, and Xinhua Dai.
\newblock A transformer-based semi-autoregressive framework for high-speed and accurate de novo peptide sequencing.
\newblock {\em Communications Biology}, 8(1):234, February 2025.

\bibitem{zhang_-primenovo_2025}
Xiang Zhang, Tianze Ling, Zhi Jin, Sheng Xu, Zhiqiang Gao, Boyan Sun, Zijie Qiu, Jiaqi Wei, Nanqing Dong, Guangshuai Wang, Guibin Wang, Leyuan Li, Muhammad Abdul-Mageed, Laks V.~S. Lakshmanan, Fuchu He, Wanli Ouyang, Cheng Chang, and Siqi Sun.
\newblock $\pi$-{PrimeNovo}: an accurate and efficient non-autoregressive deep learning model for de novo peptide sequencing.
\newblock {\em Nature Communications}, 16(1):267, January 2025.

\bibitem{lee_bidirectional_2024}
Sangjeong Lee and Hyunwoo Kim.
\newblock Bidirectional de novo peptide sequencing using a transformer model.
\newblock {\em PLOS Computational Biology}, 20(2):e1011892, February 2024.
\newblock Publisher: Public Library of Science.

\bibitem{yang_introducing_2024}
Tingpeng Yang, Tianze Ling, Boyan Sun, Zhendong Liang, Fan Xu, Xiansong Huang, Linhai Xie, Yonghong He, Leyuan Li, Fuchu He, Yu~Wang, and Cheng Chang.
\newblock Introducing $\pi$-{HelixNovo} for practical large-scale de novo peptide sequencing.
\newblock {\em Briefings in Bioinformatics}, 25(2):bbae021, March 2024.

\bibitem{eloff_instanovo_2025}
Kevin Eloff, Konstantinos Kalogeropoulos, Amandla Mabona, Oliver Morell, Rachel Catzel, Esperanza Rivera-de Torre, Jakob Berg~Jespersen, Wesley Williams, Sam P.~B. van Beljouw, Marcin~J. Skwark, Andreas~Hougaard Laustsen, Stan J.~J. Brouns, Anne Ljungars, Erwin~M. Schoof, Jeroen Van~Goey, Ulrich auf~dem Keller, Karim Beguir, Nicolas Lopez~Carranza, and Timothy~P. Jenkins.
\newblock {InstaNovo} enables diffusion-powered de novo peptide sequencing in large-scale proteomics experiments.
\newblock {\em Nature Machine Intelligence}, 7(4):565--579, April 2025.

\bibitem{tran_novoboard_2024}
Ngoc~Hieu Tran, Rui Qiao, Zeping Mao, Shengying Pan, Qing Zhang, Wenting Li, Lei Xin, Ming Li, and Baozhen Shan.
\newblock {NovoBoard}: {A} {Comprehensive} {Framework} for {Evaluating} the {False} {Discovery} {Rate} and {Accuracy} of {De} {Novo} {Peptide} {Sequencing}.
\newblock {\em Molecular \& Cellular Proteomics}, 23(11), November 2024.
\newblock Publisher: Elsevier.

\bibitem{qiu2025universalbiologicalsequencereranking}
Zijie Qiu, Jiaqi Wei, Xiang Zhang, Sheng Xu, Kai Zou, Zhi Jin, Zhiqiang Gao, Nanqing Dong, and Siqi Sun.
\newblock Universal biological sequence reranking for improved de novo peptide sequencing, 2025.

\bibitem{yang_msbooster_2023}
Kevin~L. Yang, Fengchao Yu, Guo~Ci Teo, Kai Li, Vadim Demichev, Markus Ralser, and Alexey~I. Nesvizhskii.
\newblock {MSBooster}: improving peptide identification rates using deep learning-based features.
\newblock {\em Nature Communications}, 14(1):4539, July 2023.

\bibitem{kalhor_rescoring_2024}
Mostafa Kalhor, Joel Lapin, Mario Picciani, and Mathias Wilhelm.
\newblock Rescoring {Peptide} {Spectrum} {Matches}: {Boosting} {Proteomics} {Performance} by {Integrating} {Peptide} {Property} {Predictors} {Into} {Peptide} {Identification}.
\newblock {\em Molecular \& Cellular Proteomics}, 23(7):100798, 2024.

\bibitem{miller_postnovo_2018}
Samuel~E. Miller, Adriana~I. Rizzo, and Jacob~R. Waldbauer.
\newblock Postnovo: {Postprocessing} {Enables} {Accurate} and {FDR}-{Controlled} de {Novo} {Peptide} {Sequencing}.
\newblock {\em Journal of Proteome Research}, 17(11):3671--3680, November 2018.
\newblock Publisher: American Chemical Society.

\bibitem{sanders_transformer_2024}
Justin Sanders, Bo~Wen, Paul Rudnick, Rich Johnson, Christine~C. Wu, Sewoong Oh, Michael~J. MacCoss, and William~Stafford Noble.
\newblock A transformer model for de novo sequencing of data-independent acquisition mass spectrometry data.
\newblock {\em bioRxiv}, 2024.

\bibitem{sanders_procedure_2025}
Justin Sanders, William~Stafford Noble, and Uri Keich.
\newblock A procedure for controlling the false discovery rate of de novo peptide sequencing.
\newblock {\em bioRxiv}, 2025.
\newblock Publisher: Cold Spring Harbor Laboratory \_eprint: https://www.biorxiv.org/content/early/2025/09/17/2025.09.12.675837.full.pdf.

\bibitem{gessulat_prosit_2019}
Siegfried Gessulat, Tobias Schmidt, Daniel~Paul Zolg, Patroklos Samaras, Karsten Schnatbaum, Johannes Zerweck, Tobias Knaute, Julia Rechenberger, Bernard Delanghe, Andreas Huhmer, Ulf Reimer, Hans-Christian Ehrlich, Stephan Aiche, Bernhard Kuster, and Mathias Wilhelm.
\newblock Prosit: proteome-wide prediction of peptide tandem mass spectra by deep learning.
\newblock {\em Nature Methods}, 16(6):509--518, June 2019.

\bibitem{escher_using_2012}
Claudia Escher, Lukas Reiter, Brendan MacLean, Reto Ossola, Franz Herzog, John Chilton, Michael~J. MacCoss, and Oliver Rinner.
\newblock Using {iRT}, a normalized retention time for more targeted measurement of peptides.
\newblock {\em Proteomics}, 12(8):1111--1121, April 2012.
\newblock Place: Germany.

\bibitem{perezriverol_pride_2021}
Yasset Perez-Riverol, Jingwen Bai, Chakradhar Bandla, David García-Seisdedos, Suresh Hewapathirana, Selvakumar Kamatchinathan, Deepti J Kundu, Ananth Prakash, Anika Frericks-Zipper, Martin Eisenacher, Mathias Walzer, Shengbo Wang, Alvis Brazma, and Juan Antonio Vizcaíno.
\newblock The pride database resources in 2022: a hub for mass spectrometry-based proteomics evidences.
\newblock {\em Nucleic Acids Research}, 50(D1):D543--D552, 11 2021.

\bibitem{beslic_comprehensive_2023}
Denis Beslic, Georg Tscheuschner, Bernhard~Y Renard, Michael~G Weller, and Thilo Muth.
\newblock Comprehensive evaluation of peptide de novo sequencing tools for monoclonal antibody assembly.
\newblock {\em Brief. Bioinform.}, 24(1), January 2023.

\bibitem{nguyen_highthroughput_2022}
Giang Thi~Tuyet Nguyen, Carol O'Brien, Yessica Wouters, Lorenzo Seneci, Alex Gallissà-Calzado, Isabel Campos-Pinto, Shirin Ahmadi, Andreas~H Laustsen, and Anne Ljungars.
\newblock High-throughput proteomics and in vitro functional characterization of the 26 medically most important elapids and vipers from sub-saharan africa.
\newblock {\em GigaScience}, 11:giac121, 12 2022.

\bibitem{mikosinski_longitudinal_2022}
Jacek Mikosi{\'n}ski, Konstantinos Kalogeropoulos, Louise Bundgaard, Cathrine~Agnete Larsen, Simonas Savickas, Aleksander Moldt~Haack, Konrad Pa{\'n}czak, Katarzyna Rybo{\l}owicz, Tomasz Grzela, Micha{\l} Olszewski, Piotr Ciszewski, Karina Sitek-Zi{\'o}{\l}kowska, Krystyna Twardowska-Saucha, Marek Karczewski, Daniel Rabczenko, Agnieszka Segiet, Patrycja Buczak-Kula, Erwin~M Schoof, Sabine~A Eming, Hans Smola, and Ulrich Auf~dem Keller.
\newblock Longitudinal evaluation of biomarkers in wound fluids from venous leg ulcers and split-thickness skin graft donor site wounds treated with a protease-modulating wound dressing.
\newblock {\em Acta Derm. Venereol.}, 102:adv00834, December 2022.

\bibitem{muller_proteome_2020}
Johannes~B. Müller, Philipp~E. Geyer, Ana~R. Colaço, Peter~V. Treit, Maximilian~T. Strauss, Mario Oroshi, Sophia Doll, Sebastian Virreira~Winter, Jakob~M. Bader, Niklas Köhler, Fabian Theis, Alberto Santos, and Matthias Mann.
\newblock The proteome landscape of the kingdoms of life.
\newblock {\em Nature}, 582(7813):592--596, June 2020.

\bibitem{chong_highthroughput_2018}
Chloe Chong, Fabio Marino, Huisong Pak, Julien Racle, Roy~T Daniel, Markus M{\"u}ller, David Gfeller, George Coukos, and Michal Bassani-Sternberg.
\newblock High-throughput and sensitive immunopeptidomics platform reveals profound interferon$\gamma$-mediated remodeling of the human leukocyte antigen ({HLA}) ligandome.
\newblock {\em Mol. Cell. Proteomics}, 17(3):533--548, March 2018.

\end{thebibliography}

\section{Acknowledgements}
K.K. is supported by a Novo Nordisk Foundation Young Investigator Award (grant no. NNF16OC0020670) and a postdoctoral fellowship grant from the Independent Research Fund Denmark (grant no. 4257-00010B). 
We are grateful to the DTU Proteomics Core Facility for maintenance and running of mass spectrometry instruments. 
We also thank the entire InstaNovo team for their valuable input and feedback during this study.

\section{Author contributions statement}

K.K. and A.M. conceived the project. 
K.K. provided datasets for validation. 
A.M. and J.D. preprocessed the data, wrote the software and trained models with feedback from K.E., E.M.S, T.P.J. and K.K.
A.M., J.D., H.S.J.K. and K.K. analysed the output and performed benchmarking with feedback from R.C., K.E., E.M.S. and J.V.G.
K.K., N.L.C., T.P.J. and J.V.G. supervised the project. 
A.M., J.D. and K.K. drafted the original manuscript. 
All authors reviewed the manuscript and approved its final version.

\section{Declaration of generative AI and AI-assisted technologies in the writing process}

During the preparation of this work, the authors used GPT-4o in order to improve language and readability.
After using this tool, the authors reviewed and edited the content as needed and take full responsibility for the content of the manuscript.

\section{Additional information}

\renewcommand{\figurename}{Supplementary Figure}
\renewcommand{\tablename}{Supplementary Table}
\setcounter{figure}{0}

\subsection{Declaration of competing interests} 
R.C, J.D, A.M., K.E, N.L.C. and J.V.G. are employees of InstaDeep, 5 Merchant Square, London, UK. The other authors declare no competing interests.

\newpage

\subsection{Supplementary tables}

\begin{table}[h]
    \centering
    \caption{\textbf{Performance metrics on the HeLa Single Shot dataset at 5\% FDR, using a calibrator trained on HeLa Single Shot data.} We use correct proteome hit as a proxy for correct sequence label in the full search space and PSM correctness via database search in the labelled space. Winnow calibration improves recall on the labelled set while controlling FDR accurately, and maintains FDR control with reduced recall in the full search space compared to the existing database-grounded method.}
    \label{tab:hela_results}
    \begin{tabular}{llcccccc}
        \toprule
        Dataset & Confidence \& FDR Method & Confidence Cutoff & Recall & FDR \\
        \midrule
        \multirow{2}{*}{HeLa Single Shot (Labelled)}
        & Calibrated, Winnow              & 0.854 & 0.741 & 0.046 \\
        & Raw, Database-grounded          & 0.576 & 0.660 & 0.050 \\
        \midrule
        \multirow{2}{*}{HeLa Single Shot (Full)}
        & Calibrated, Winnow              & 0.916 & 0.157 & 0.046 \\
        & Raw, Database-grounded          & 0.576 & 0.489 & 0.073 \\
        \bottomrule
    \end{tabular}
\end{table}

\begin{table}[h]
    \centering
    \caption{\textbf{Performance metrics across external datasets at 5\% FDR, using the pre-trained general calibrator.} We use correct proteome hit as a proxy for correct sequence label in the full search space and PSM correctness via database search in the labelled space. For \textit{C. elegans} datasets, calibrated Winnow improves recall over raw confidence while controlling FDR similarly to the current DNS FDR control approach. For Immunopeptidomics-2, database-grounded calibration achieves higher recall but with higher FDR, whereas calibrated Winnow maintains stricter FDR control at the cost of lower recall.}
    \label{tab:external_results}
    \begin{tabular}{llcccccc}
        \toprule
        Dataset & Confidence \& FDR Method & Confidence Cutoff & Recall & FDR \\
        \midrule
        \multirow{2}{*}{\textit{C. elegans} (Labelled)}
        & Calibrated, Winnow              & 0.913 & 0.236 & 0.042 \\
        & Raw, Database-grounded          & 0.954 & 0.089 & 0.050 \\
        \midrule
        \multirow{2}{*}{\textit{C. elegans} (Full)}
        & Calibrated, Winnow              & 0.914 & 0.174 & 0.077 \\
        & Raw, Database-grounded          & 0.954 & 0.060 & 0.076 \\
        \midrule
        \multirow{2}{*}{Immunopeptidomics-2 (Labelled)}
        & Calibrated, Winnow              & 0.932 & 0.024 & 0.000 \\
        & Raw, Database-grounded          & 0.536 & 0.066 & 0.047 \\
        \midrule
        \multirow{2}{*}{Immunopeptidomics-2 (Full)}
        & Calibrated, Winnow              & 0.931 & 0.016 & 0.076 \\
        & Raw, Database-grounded          & 0.536 & 0.042 & 0.091 \\
        \bottomrule
    \end{tabular}
\end{table}

\subsection{Supplementary figures}

\begin{figure}
    \centering
    \includegraphics[width=0.8\linewidth]{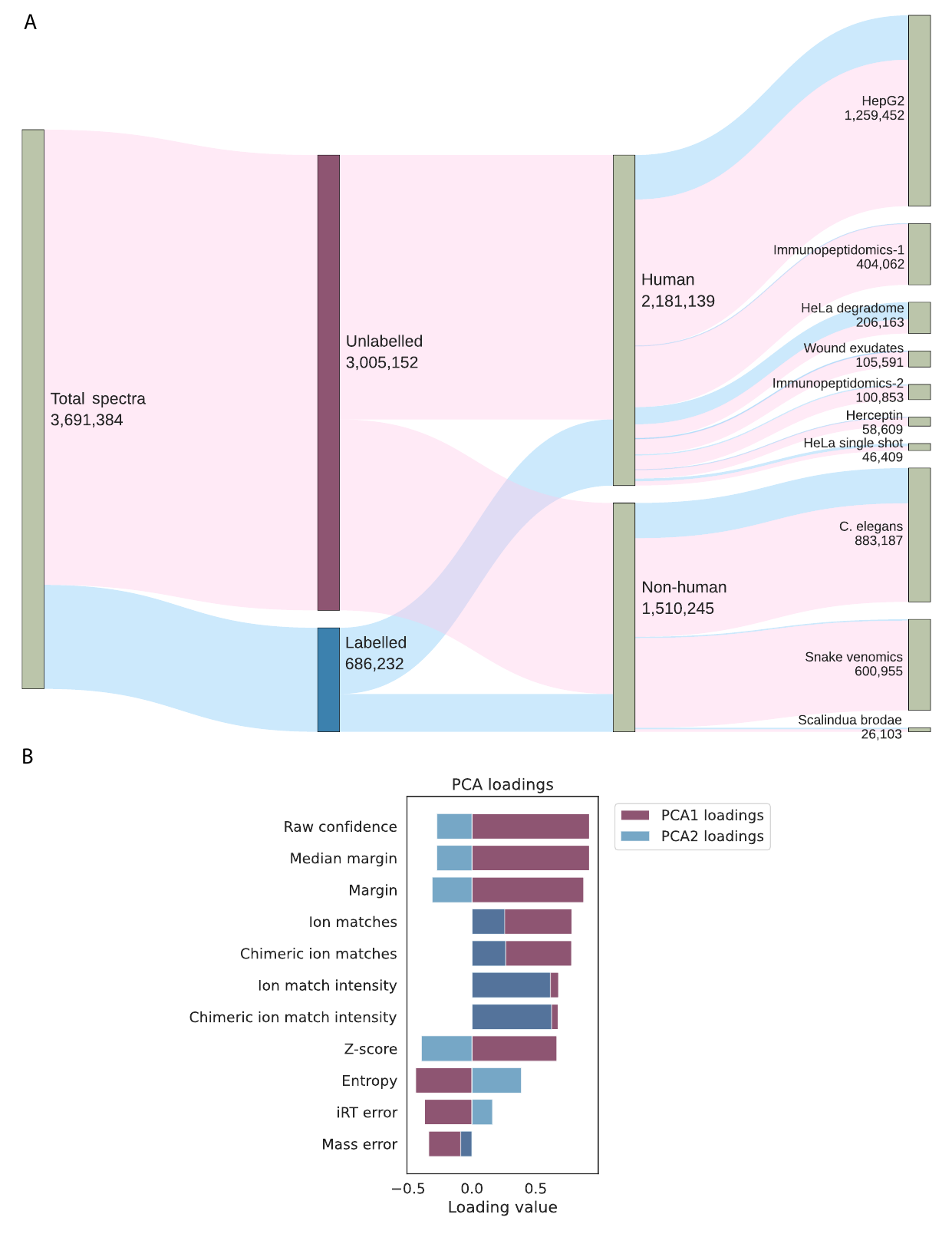}
    \caption{\textbf{Dataset composition and exploratory analysis.} \textbf{A}) Sankey plot showing the composition of the general calibrator model’s training and test data, illustrating the contribution of different dataset sources, the proportion of spectra labelled by database search, and the proportion of human and non-human data. \textbf{B}) Principal component analysis (PCA) of PSM features from the HeLa Single Shot test dataset, with the first two components and their loadings shown, highlighting the features driving variance in the data. Features are ordered by descending absolute loading value for the first principle component.}
    \label{fig:dataset_characteristics}
\end{figure}

\begin{figure}
    \centering
    \includegraphics[width=1\linewidth]{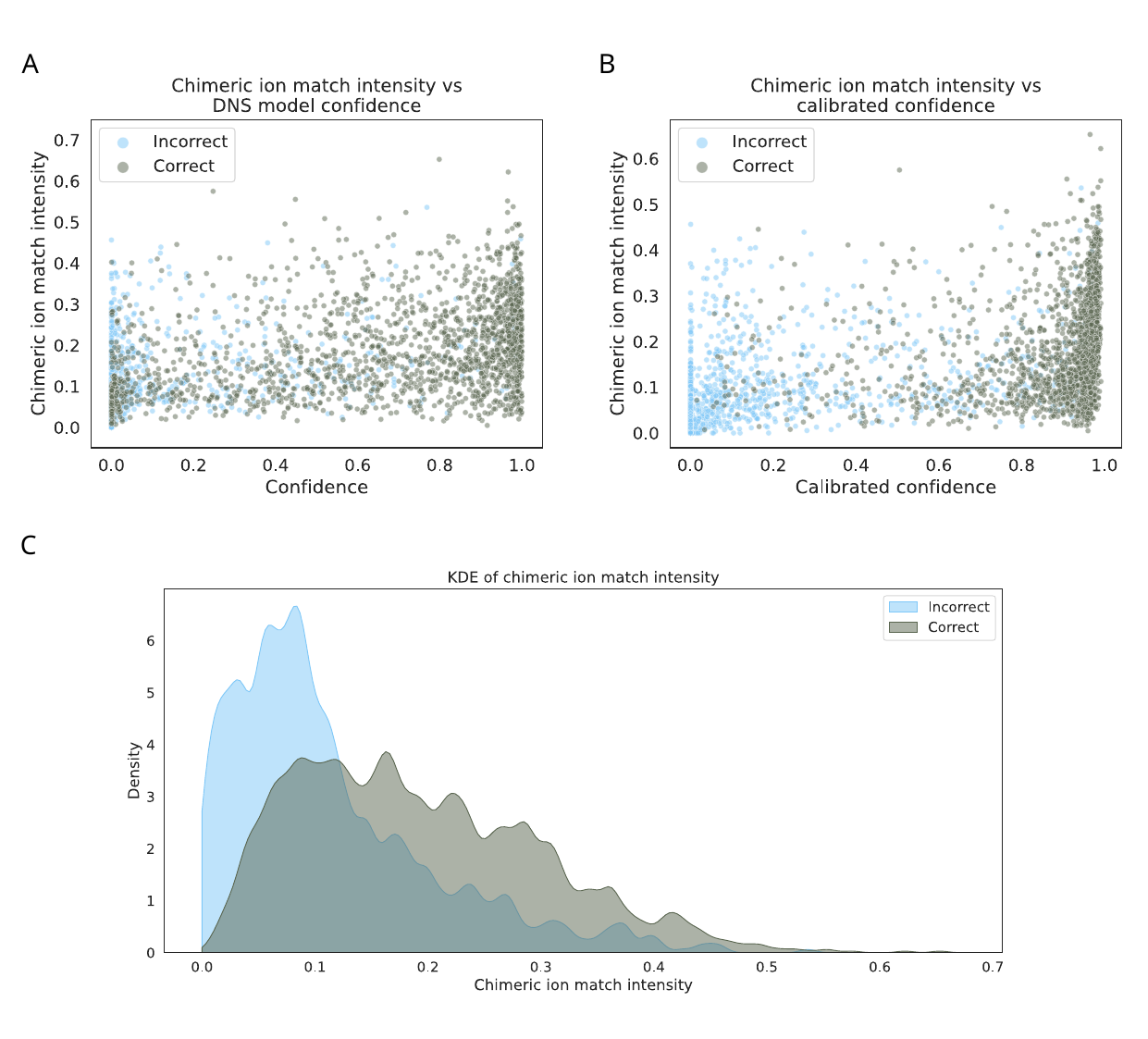}
    \caption{\textbf{Chimeric ion match intensity feature analysis using the labelled HeLa Single Shot test dataset.} \textbf{A}) Relationship between chimeric ion match intensity and raw DNS model confidence. \textbf{B}) Relationship between chimeric ion match intensity and calibrated confidence. \textbf{C}) Kernel-density estimate (KDE) plot of chimeric ion match intensity, showing the distribution of correct and incorrect PSM identifications according to database search results.}
    \label{fig:chimeric}
\end{figure}

\begin{figure}
    \centering
    \includegraphics[width=0.9\linewidth]{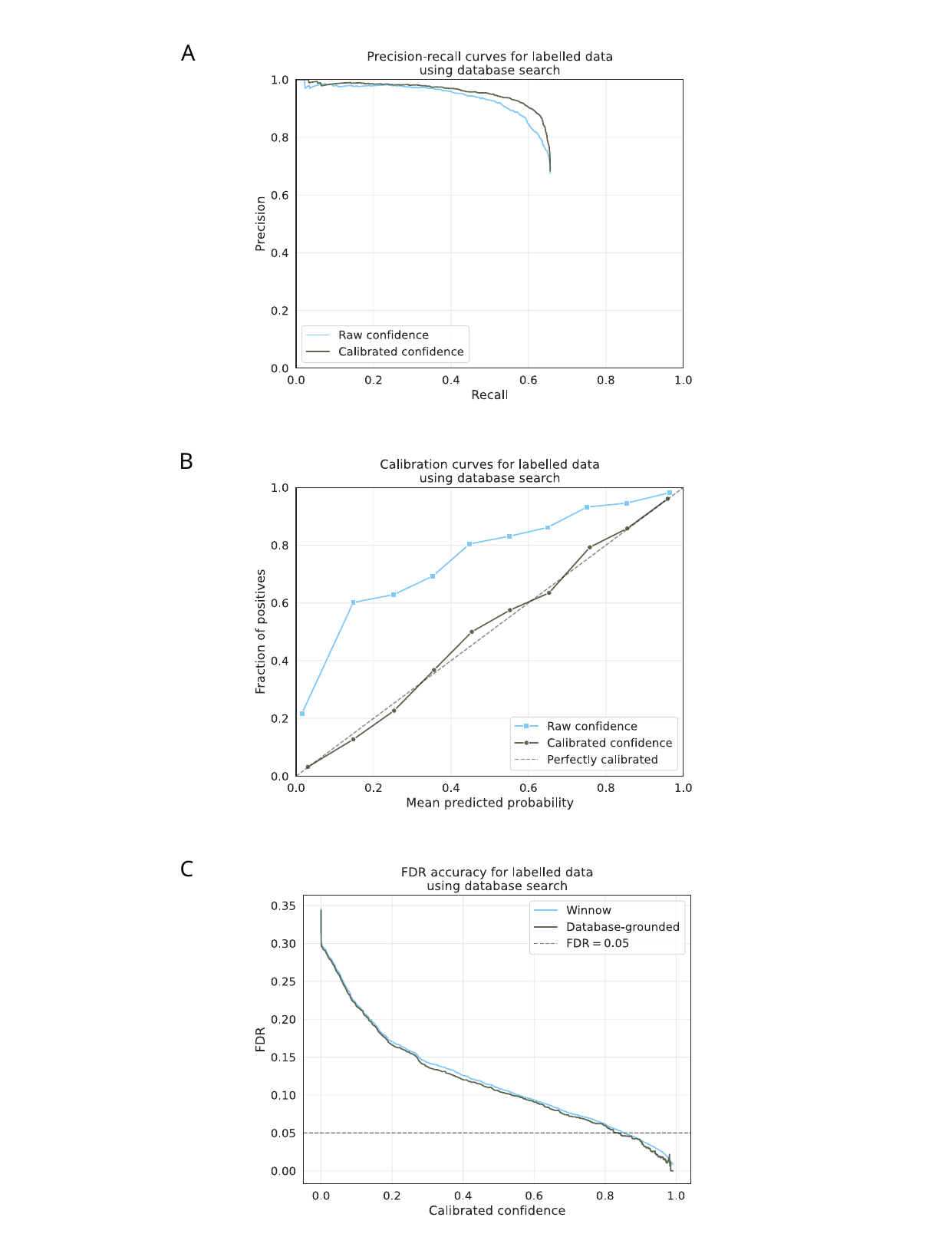}
    \caption{\textbf{Performance of Winnow’s full pipeline trained and evaluated on portions of the HeLa Single Shot dataset.} A) Precision-recall curves comparing calibrated and raw DNS model confidence on the labelled test set of HeLa Single Shot. B) Calibration curves for calibrated and raw confidence, compared against perfect calibration, for the labelled test set. C) PSM-specific FDR run plots for Winnow’s non-parametric and database-grounded FDR estimation methods, using calibrated confidence, for the labelled test set.}
    \label{fig:hela_test_with_database_search}
\end{figure}

\begin{figure}
    \centering
    \includegraphics[width=0.9\linewidth]{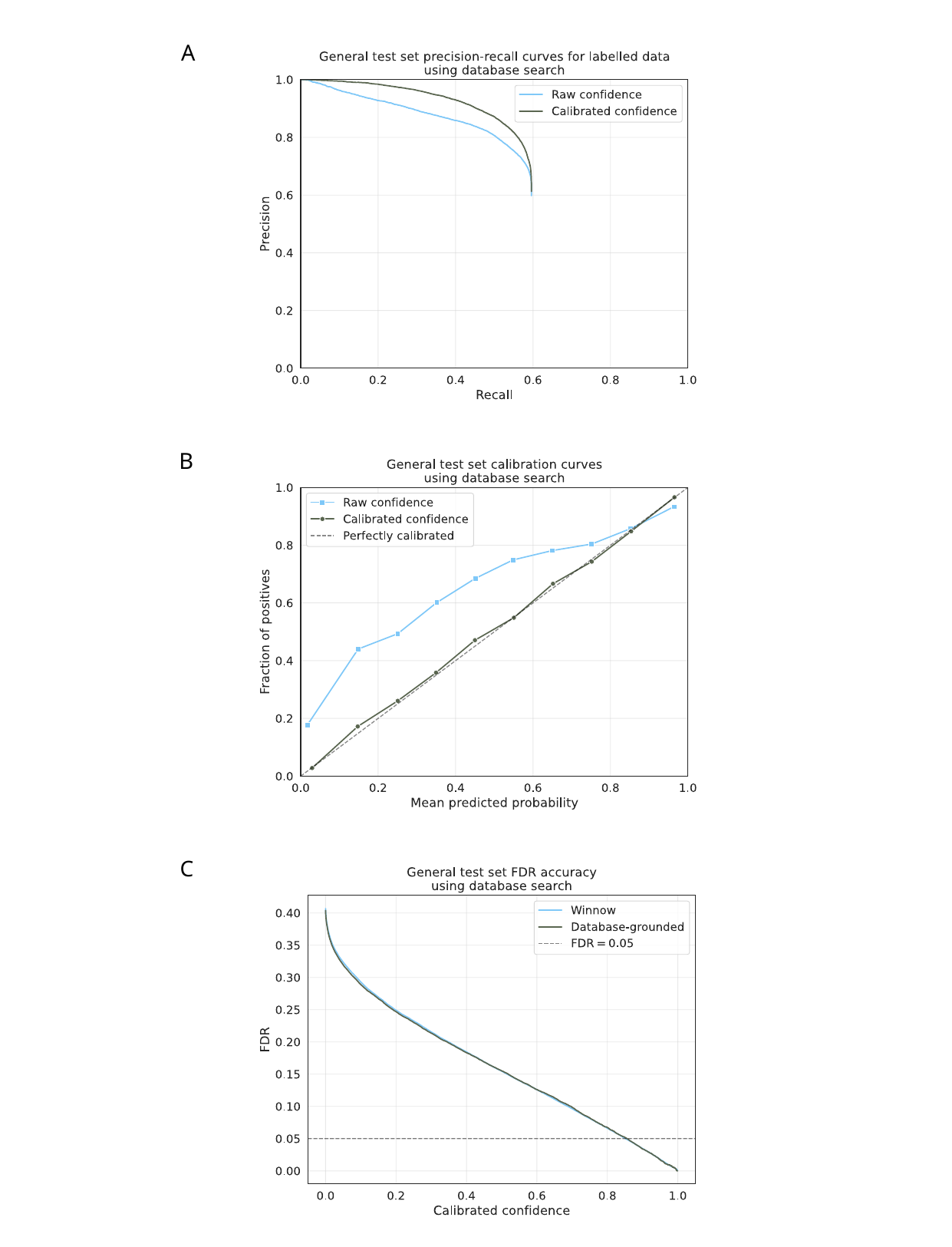}
    \caption{\textbf{Performance of Winnow's full pipeline on the general model test set, labelled using database search.} \textbf{A}) Precision-recall curves comparing calibrated and raw DNS model confidence. \textbf{B}) Calibration curves for calibrated and raw confidence, compared against perfect calibration. \textbf{C}) PSM-specific FDR run plots for Winnow's non-parametric and database-grounded FDR estimation methods using calibrated confidence.}
    \label{fig:general_test_performance}
\end{figure}

\begin{figure}
    \centering
    \includegraphics[width=1\linewidth]{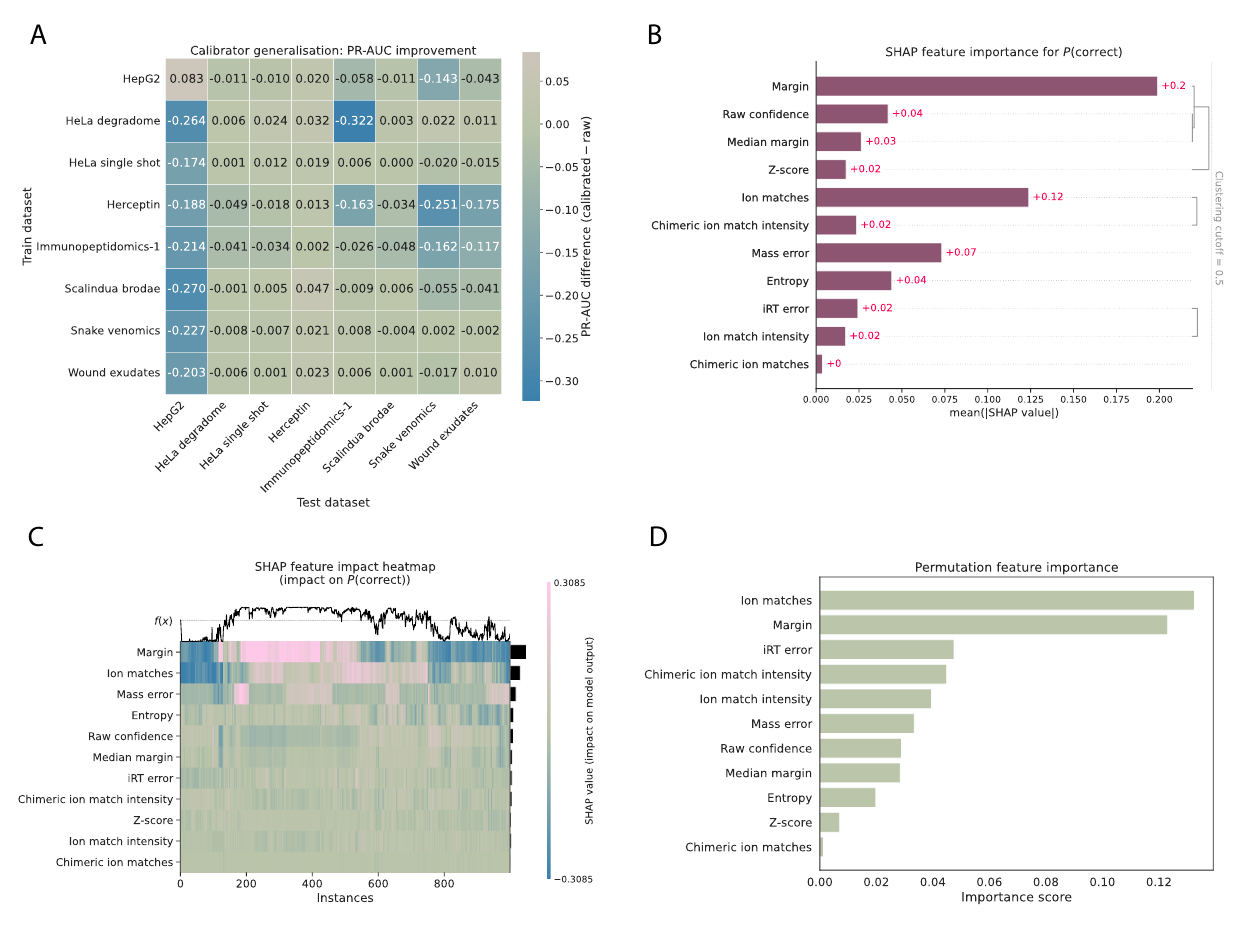}
    \caption{\textbf{Calibrator performance improvement over raw DNS model confidence and feature contributions during calibration.} \textbf{A}) Improvement in precision-recall AUC (PR-AUC) from calibration relative to raw DNS model confidence, shown in a leave-one-out dataset analysis across the general model’s training datasets. \textbf{B}) SHAP feature importance scores for the general model, grouped by feature clusters created with an XGBoost model, with a cutoff for distances less than than 0.5. Distance in the clustering is assumed to be scaled roughly between 0 and 1, where 0 distance means the features perfectly redundant and 1 means they are completely independent. \textbf{C}) SHAP heatmap showing the distribution of feature impacts across individual predictions. Each row represents a single prediction, clustered to group similar feature contributions, while each column corresponds to a feature. Cell colour indicates the SHAP value, showing whether a feature increases (pink) or decreases (blue) calibrated confidence for that sample. The line above the heatmap depicts calibrator output $f(x)$ for each row, and the bar plot on the right-hand side shows the overall SHAP-based importance ranking of each feature. Features with consistent colouring across rows have a uniform effect on predictions, whereas columns with mixed colours indicate context-dependent contributions. \textbf{D}) Feature permutation importance scores, shown as a bar plot. These are computed for each feature over ten runs by measuring the drop in calibrator performance when replaced by another randomly selected feature.}
    \label{fig:extra_feature_importance_plots}
\end{figure}

\begin{figure}
    \centering
    \includegraphics[width=1\linewidth]{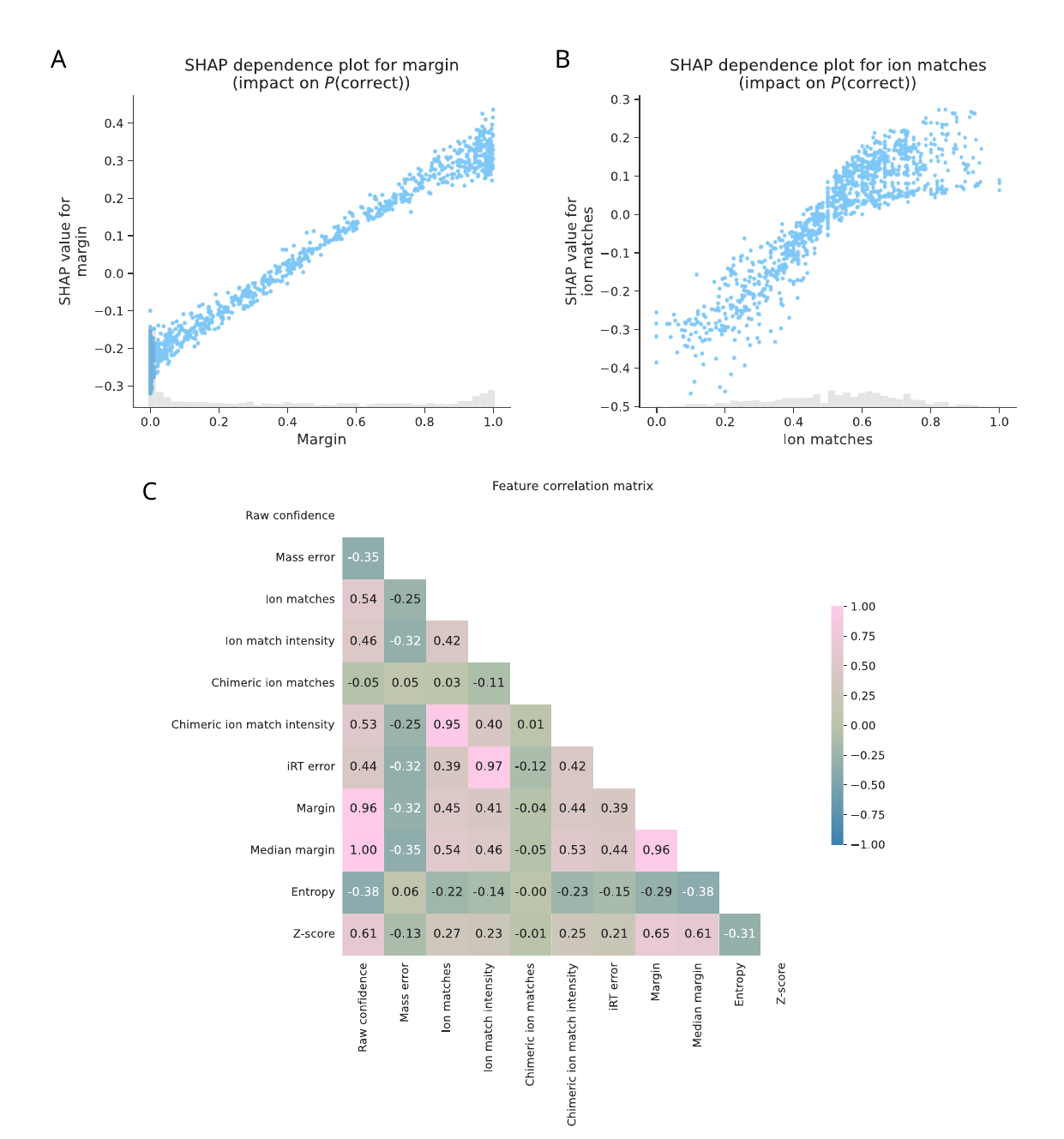}
    \caption{\textbf{Key feature effects in the general model and dataset correlations.} \textbf{A}) SHAP dependence plot for margin, the most influential feature arising from SHAP analysis and the secondmost important feature from permutation feature importance, with the underlying feature distribution shown in grey. \textbf{B}) SHAP dependence plot for ion matches, the second most influential feature from SHAP analysis and the most important feature from permutation feature importance analysis, again with the feature distribution shown in grey. \textbf{C}) Pairwise correlation matrix of calibrator input features, highlighting redundancy and complementarity, using the general model training data.}
    \label{fig:top_feature_importances}
\end{figure}

\begin{figure}
    \centering
    \includegraphics[width=1\linewidth]{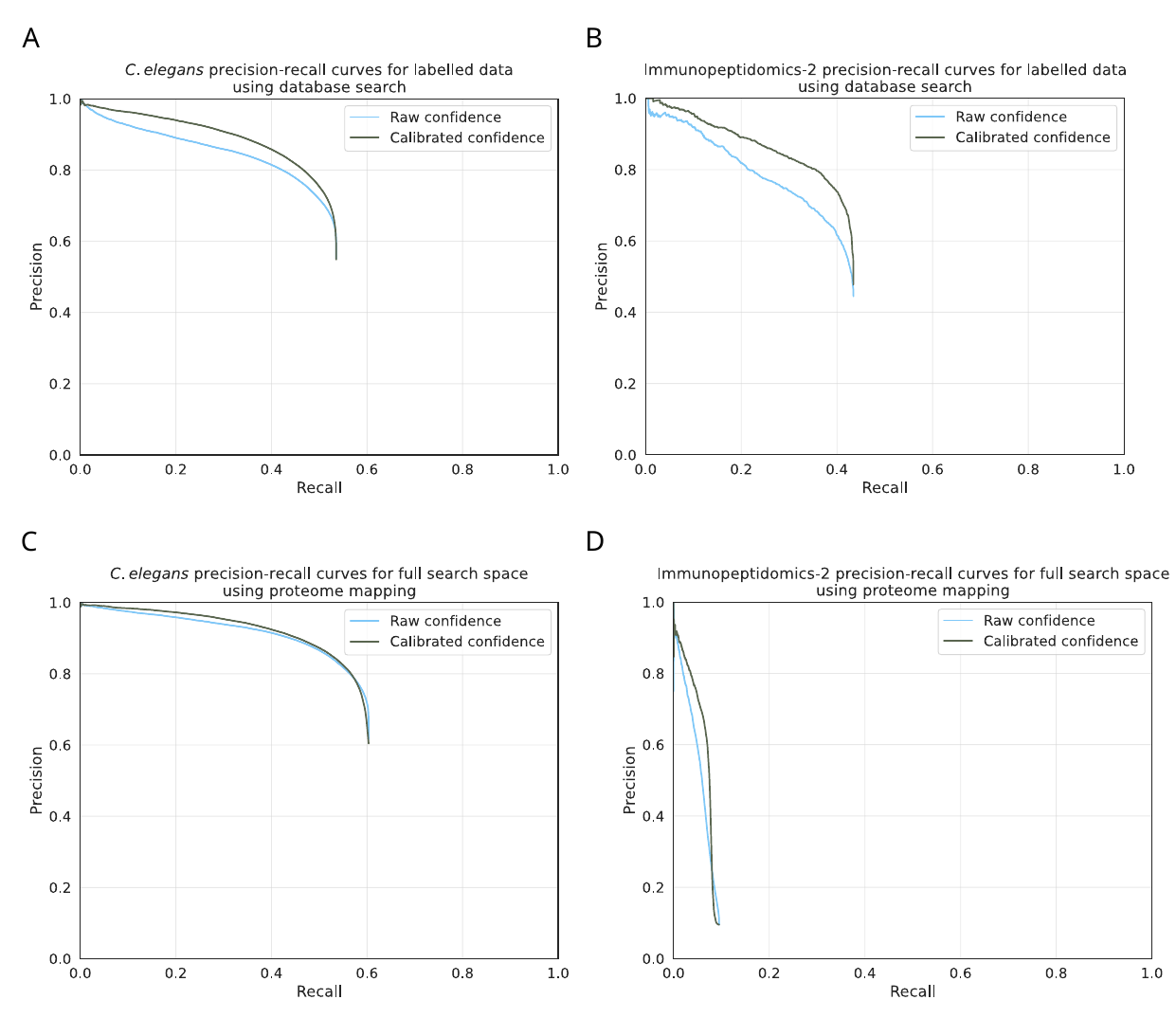}
    \caption{\textbf{General model precision-recall curves for two held-out datasets using: \textit{C. elegans} and Immunopeptidomics-2}. \textbf{A}) Precision-recall curves for the subset of \textit{C. elegans} that received database search labels, comparing raw DNS model confidence and calibrated confidence. \textbf{B}) Precision-recall curves for the labelled subset of Immunopeptidomics-2 using database search. \textbf{C}) Precision-recall curves for the full \textit{C. elegans} dataset using correct proteome mapping as a proxy for correct PSM identification. \textbf{D}) Precision-recall curves for the full Immunopeptidomics-2 dataset using correct proteome mapping as a proxy for correct PSM identification.}
    \label{fig:external_data_pr_curves}
\end{figure}

\end{document}